\documentclass[fleqn,usenatbib]{mnras}

\usepackage{newtxtext,newtxmath}

\usepackage[T1]{fontenc}
\usepackage{ae,aecompl}

\usepackage{graphicx}	
\usepackage{amsmath}	
\usepackage{color} 
\usepackage{rotating}
\usepackage{lscape}
\usepackage{subfig}
\usepackage[normalem]{ulem} 
\usepackage{subfig}
\usepackage{tablefootnote}

\def\ms{\mbox{$M_{\ast}$}}
\def\msun{\mbox{M$_{\odot}$}} 
\def\type{\mbox{\textsc{T-type}}}

\def\age{\mbox{$Age_{\rm lw}$}} 

\defcitealias{Nair2010}{NA10}
\defcitealias{DominguezSanchez2021}{DS21}

\definecolor{applegreen}{rgb}{0.55, 0.71, 0.0}

\title[MaNGA Morphology]{SDSS IV MaNGA: Visual Morphological and Statistical Characterization of the DR15 sample.}

\author[J. A. V\'azquez-Mata et al.]{
J. A. V{\'a}zquez-Mata,$^{1,2}$\thanks{E-mail: jvazquez@astro.unam.mx (KTS)}
H. M. Hern\'andez-Toledo,$^{1}$
V. Avila-Reese,$^{1}$
M. Herrera-Endoqui,$^{4}$  \and 
A. Rodr\'iguez-Puebla,$^{1}$
M. Cano-D\'iaz,$^{3}$
I. Lacerna,$^{5,6}$
L. A. Mart\'inez-V{\'a}zquez,$^{1}$
R. Lane$^{7}$
\\
\\
$^{1}$Universidad Nacional Aut\'onoma de M\'exico, Instituto de Astronom\'ia, A.P. 70-264, CDMX, 04510,  M\'exico\\
$^{2}$Departamento de F\'isica, Facultad de Ciencias, Universidad Nacional Aut\'onoma de M\'exico, Ciudad Universitaria, CDMX, 04510, M\'exico\\
$^{3}$CONACYT Research Fellow - Universidad Nacional Aut\'onoma de M\'exico, Instituto de Astronom\'ia, A.P. 70-264, CDMX, 04510,  M\'exico\\
$^{4}$Instituto de Astronom\'ia sede Ensenada, Universidad Nacional Aut\'onoma de M\'exico, Km 107, Carret. Tij.-Ens., Ensenada, 22060, BC, M\'exico\\
$^{5}$Instituto de Astronom\'ia y Ciencias Planetarias, Universidad de Atacama, Copayapu 485, Copiap\'o, Chile\\
$^{6}$Millennium Institute of Astrophysics, Nuncio Monsenor Sotero Sanz 100, Of. 104, Providencia, Santiago, Chile\\
$^{7}$Centro de Investigación en Astronomía, Universidad Bernardo O'Higgins, Avenida Viel 1497, Santiago, Chile
}

\date{Accepted XXX. Received YYY; in original form ZZZ}

\pubyear{2022}

\begin{document}
\label{firstpage}
\pagerange{\pageref{firstpage}--\pageref{lastpage}}
\maketitle


\begin{abstract}

We present a detailed visual morphological classification for the 4614 MaNGA galaxies in SDSS Data Release 15, using image mosaics generated from a combination of $r-$band (SDSS and deeper DESI Legacy Surveys) images and their digital post-processing. We distinguish 13 Hubble types and identify the presence of bars and bright tidal debris. After correcting the MaNGA sample for volume completeness, we calculate the morphological fractions, the bi-variate distribution of type and stellar mass \ms --where we recognise a morphological transition ''valley'' around S0a-Sa types-- and the variations of the $g-i$ colour and luminosity-weighted age over this distribution.
We identified bars in 46.8\% of galaxies, present in all Hubble types later than S0. This fraction amounts to a factor $\sim2$ larger when compared with other works for samples in common. We detected 14\% of galaxies with tidal features, with the fraction changing with \ms\ and morphology. For 355 galaxies, the classification was uncertain; they are visually faint, mostly of low/intermediate masses, low concentrations, and disky in nature. Our morphological classification agrees well with other works for samples in common, though some particular differences emerge, showing that our image procedures allow us to identify a wealth of added value information as compared to SDSS-based previous estimates. Based on our classification, we also propose an alternative criteria for the E--S0 separation, in the structural semi-major to semi-minor axis versus bulge to total light ratio ($b/a-B/T$) and concentration versus semi-major to semi-minor axis ($C-b/a$) space.

\end{abstract}

\begin{keywords}
Extragalactic -- morphology -- statistics
\end{keywords}



\section{Introduction}

The visual structural appearance of galaxies in optical bands, called morphology, was the 
first property studied soon after their discovery as extragalactic objects by E. Hubble.
As a result, Hubble introduced the morphological classification for luminous and local, $z\sim 0$, galaxies \citep[][]{Hubble1926,Hubble1936} named after him as the Hubble sequence.  With increasingly powerful telescopes, new structural details were revealed in galaxies, leading to more details in the Hubble classification \citep[e.g.,][]{vandenBergh1960a,vandenBergh1960b,Buta1995}, and lower luminosity galaxies were detected, which required extending this classification at its late-type end \citep[for historical reviews, see e.g.,][and references therein]{Sandage2005,Graham2019}. 
So far, the modified Hubble sequence remains the most common and useful system for the morphological classification of galaxies. 

In the most general way, the Hubble morphological sequence ranges from early types (elliptical and lenticular galaxies) to late types (spiral and irregular galaxies). The main global physical and environmental properties of galaxies change throughout this sequence, although it is still a matter of debate which of the properties are most relevant  to shaping the sequence.
From the observational point of view, structural and kinematic properties seem to be closely related to this sequence. For example, it is remarkable the increase of the bulge-to-disk ratio \citep[][]{Hubble1936,Simien+1986} and the decrease of the specific angular momentum \citep[e.g.,][]{Sandage1970,Cappellari+2011,Romanowsky+2012} when moving towards earlier types along the sequence. Also, intensive properties related to the star formation history of galaxies such as colour and specific star formation rate (sSFR), and in a lesser extent, extensive properties as luminosity and stellar mass (M$_{*}$), are correlated with the morphological type  \citep[e.g.,][]{Roberts+1994,Blanton2009}.  It is also well known that earlier-type galaxies are more frequent in denser environments \citep[e.g.,][]{Dressler1980, Park2007, Pearson2021}, while in the low-density environments, most of galaxies are of late types.

Within the current theoretical scenario of galaxy formation and evolution \citep[see e.g.,][]{Mo+2010}, galaxies emerge from the inflow and cooling of cosmic gas driven, in first approximation, by the mass accretion rate of the dark matter halos, where they are formed. As a result, several of the main properties of galaxies are expected to be imprinted by the properties and mass accretion rate of the haloes \citep[e.g.,][]{Avila-Reese_Firmani2000,Bouche+2010,Dave+2012,Dekel+2013,Rodriguez-Puebla+2016}. The haloes grow not only by smooth accretion but also by mergers, which ultimately may lead to strong interactions and the merging of galaxies residing  in their centers. The outcome of galaxy mergers, specially major mergers, is the destruction of the disks in favor of dynamically hot spheroids. Thus, the interplay between cosmic cool gas accretion onto galaxies and galaxy mergers, as well as star formation and Active Galactic Nuclei (AGN) formation and their feedback effects, and dynamical secular processes,  give rise to the observed morphological diversity of galaxies as state-of-the-art hydrodynamics cosmological simulations have shown  \citep[e.g.,][]{Vogelsberger2014,Schaye+2015,Dubois2016}. In this context, a reliable morphological classification becomes crucial for understanding the properties and formation mechanisms of galaxies, and for constraining models and simulations of galaxy evolution.

Since the Hubble morphological classification is based on the recognition of structural features seen on high quality images of galaxies, their visual inspection by a group of experts, supported by an appropriate digital processing, remains as the most reliable way to determine the Hubble type of galaxies \citep[see e.g.][]{HernandezToledo2010, Nair2010}. However, this approach is not feasible for large surveys observing hundreds of thousands objects, as the Sloan Digital Sky Survey (SDSS) sample \citep{York2000}, or the new generation of extragalactic surveys, which will observe tens of millions of objects, as the Dark Energy Spectroscopy Instrument (DESI) Legacy Surveys \citep{Dey2019}, the Legacy Survey of Space and Time \citep[LSST;][]{Ivezic2019}, etc. A great effort to attain the morphological classification for the SDSS galaxies by visual methods was the Galaxy Zoo (GZ;  \citealt{Lintott2008}, \citealt{Willett2013}) "citizen science" project, in which thousands of volunteers were asked to classify galaxies' morphological features. More recently, machine and deep learning methods have been developed and applied for classifying galaxies from the SDSS and DESI Legacy surveys  \citep[e.g.,][]{Huertas-Company+2011,DominguezSanchez2018,Martin2019,Walmsley2021} and more recently from the Dark Energy Survey \citep[DES][]{Cheng2021, Vega-Ferrero2021}.

Recently, several medium-size surveys of thousands of local galaxies have been, or are in the process of being completed using the Integral Field Spectroscopic (IFS) technique, which obtain spatially resolved spectra for extended objects using arrays of integral field units (IFUs). Among these surveys are the Calar Alto Legacy Integral Field Area survey \citep[CALIFA;][]{Sanchez2012}, Sydney-AAO Multi-object IFS \citep[SAMI;][]{Croom2012}, and Mapping Nearby Galaxies at Apache Point Observatory \citep[MaNGA;][]{Bundy2015, Yang2016}. From the IFS surveys, fundamental sets of spatially-resolved spectral galaxy properties are obtained,
which by applying models of stellar population synthesis (SPS), allows us to reconstruct the histories of star formation, chemical enrichment, and stellar mass growth of galaxies \citep[for recent reviews, see e.g.,][]{Conroy2013,Sanchez2020}.

The current largest IFS survey of local galaxies is MaNGA, which is one of the main projects of the SDSS IV international collaboration \citep{Blanton2017}. MaNGA has been recently completed and it contains $\sim$ 10,000 observed galaxies. The IFU data from MaNGA were or are being analyzed with different pipeline/tools, for example, the official Data Analysis Pipeline \citep[DAP;][]{Westfall2019}, Fitting IteRativEly For Likelihood analYsis \citep[FIREFLY;][]{Goddard2017b}, and Pipe3D/FIT3D \citep[][]{Sanchez2016,Sanchez2018}. The use of Pipe3D to analyze the first half of MaNGA galaxies (SDSS Data Release 15, DR15) led to the generation of a Value Added Catalogue (VAC) that is publicly available.\footnote{\url{https://www.sdss.org/dr15/manga/manga-data/manga-pipe3d-value-added-catalog/}} The results from applying the SPS analysis techniques to IFS observations of local galaxies, in combination with a detailed knowledge of their morphologies, open up new alternatives to investigate the processes behind galaxy formation and evolution, offering the possibility of connecting the global physical and morphological properties of galaxies to their spatially-resolved properties at the kpc scales \citep[see, e.g.,][for a recent review see \citealp{Sanchez2020}, and more references therein]{Perez+2013,GonzalezDelgado2015,Ibarra-Medel+2016,Sanchez2018,CanoDiaz2019}. 
A detailed information on the morphological properties of the MaNGA galaxies, combined with the set of parameters derived from the IFS analysis, set the possibility of novel studies about the origin of the Hubble sequence to a new level \citep[e.g.,][]{GonzalezDelgado2016,GonzalezDelgado+2017,Goddard2017b,Peterken+2021a}.

Among the various attempts to morphologically classify the MaNGA galaxies, we mention two important ones.
The first makes use of results from the GZ project. The GZ team have compiled all the MaNGA targets, spread in 5 different GZ types data sets, into the MaNGA-GalaxyZoo catalogue.\footnote{\href{https://www.sdss.org/dr16/data_access/value-added-catalogs/?vac_id=manga-morphologies-from-galaxy-zoo}{MaNGA Morphologies from Galaxy Zoo}}
The result has been useful for studies requiring only a very general morphological classification, as a detailed Hubble classification is not provided by GZ. This classification is based on the recommended criteria of \citet{Willett2013} to obtain clean samples of galaxies of each morphology; however, there is a large number of objects where the classification is uncertain and where the vote fractions do not provide more definite results, due to intermediate morphology or distance. The second attempt is based on machine learning techniques \citep[][hereafter DS21]{Fischer2019, DominguezSanchez2021}, considering the \citet{DominguezSanchez2018} algorithm and using as a training input the \citet[][hereafter NA10]{Nair2010} classification (T-Types) and GZ catalogue. Although these results show high accuracy, the input classification could be biased by the quality of images used. 
At this point, it is desirable to also have an exhaustive visual classification of MaNGA galaxies
 
The size of the MaNGA survey, containing  $\sim$ 10,000 galaxies, makes feasible a {\it detailed and direct visual morphological classification}. In this paper, we present the visual morphological Hubble classification for 4614 MaNGA galaxies from the SDSS DR15 \citep{Aguado2019}, which represents about half of the total number of galaxies in the final MaNGA sample to be released in the final SDSS IV data release (DR17).
A remarkable advantage of our study is that it makes use of the recent photometric data provided by the DESI Legacy Surveys. The DESI Legacy Surveys have a broad coverage over the MaNGA DR15 sample, with deep $r-$band images having a median surface brightness limit of $\sim$ 27.5 mag arcsec$^{-2}$  \citep{Hood2018}, allowing us to  probe fainter tidal features than the nominal $r-$band SDSS limit of 24.6 mag arcsec$^{-2}$. In addition, this survey provides residual images (after subtraction of a best 2D brightness model) from a post-processing procedure that we use in our morphological analysis. 

Along with the Hubble morphological types, we visually classify other relevant morphological features of galaxies such as bars and tidal structures. 
To carry out a statistical characterization, weights were applied to transform the MaNGA sample into a volume-limited sample complete in mass above $\ms\approx 10^9$ \msun\ (as in \citealt{Sanchez+2019}, see also \citealp{Rodriguez-Puebla+2020}). From that, we are able to determine the fractions in the local universe of the different morphological types of galaxies with bars and tidal features, as well as the joint distributions of morphology with $\ms$. In forthcoming papers, we will be using the wealth of morphological information, obtained with a set of IFS parameters, to study in more detail the joint distributions of morphology with colours, sSFR, mass- and luminosity-weighted age, as well as with the radial distributions of these properties.

The content of this paper is as follows. Section~\ref{sec:manga} presents a brief summary of properties of the MaNGA sample, related to its selection criteria and IFU spatial coverage. In Section~\ref{sec:method} the procedures behind our direct visual morphological classification are described and illustrated, using our new re-processing of the SDSS images and an additional image processing of the DESI Legacy Surveys images. Section~\ref{sec:results} describes the results of our visual classification, including the identification of bars and tidal features. Section~\ref{sec:compare} presents a general comparison with a sample in common with \citetalias{DominguezSanchez2021}, \citetalias{Nair2010} and MaNGA-GalaxyZoo. In Section~\ref{sec:Discussion} we present a general discussion of our morphological results and the correlation with stellar mass, colour and age. Finally, Section~\ref{sec:conclusions} presents our final conclusions. For this work, we assume cosmological parameters of $\Omega_M$ = 0.3, $\Omega_{\Lambda}$ = 0.7 and $h$ = 0.71.

\section{The MaNGA survey}
\label{sec:manga}

The MaNGA observations make use of specially-designed IFUs of five sizes ranging from 12 to 32 arcsec diameters with 19 to 127 fibres \citep{Drory+2015}, which are mounted onto plates on the 2.5-meter Sloan telescope at APO \citep{Gunn2006}. The spectra cover a wavelength range of 3600-10300\AA. Observations are designed to reach a minimum signal-to-noise ratio of 5\AA$^{-1}$ at 1.5$R_e$ \citep{Law2015}, where $R_e$ is the $r-$band  effective radius of each observed galaxy measured by the NASA-Sloan Atlas \citep[NSA;][]{Blanton2011}.
As part of the sample selection, MaNGA targets galaxies with a flat distribution in $r-$band magnitude, $M_r$, to approximately sample a flat distribution in $\log\ms$ over the range of $10^9-10^{11.5}$ \msun\ \citep{Bundy2015}. 
The existence of well-known scaling relations for galaxies motivated the desire to have roughly uniform radial coverage as defined by some multiple of $R_e$ \citep{Wake2017}, requiring this, a stellar mass-dependent lower redshift limit, so that larger more massive galaxies have the same angular size as smaller lower mass galaxies. 
As a compromise, the full targeting catalogue assigns galaxies in the redshift range $0.01<z<0.15$ (median $z\sim 0.03$) to cover up to $1.5R_e$ for $\sim 45\%$ of the sample and up to $2.5R_e$ for $\sim 35.6\%$ of the sample. These are called the Primary and Secondary samples. 
The former is supplemented by a colour-enhanced sample ($\sim 15\%$ of the targets) to form the Primary+ sample, which over-samples unusual regions of the $(NUV - i)-M_i$ diagram such as high-mass blue galaxies, low-mass red galaxies, and the “green valley” \citep{Law2015}.
The remaining $\sim 4.4\%$ of the galaxies correspond to ancillary samples.

In this paper, we determine the morphological classification for the MaNGA galaxies reported in the SDSS DR15 \citep[][]{Aguado2019} and corresponding to the internally released MaNGA Product Launch-7 (MPL-7). 
An important source of photometric information, useful for the analysis of the morphological results in the present paper is the NSA \citep[][]{Blanton2011} catalogue. From this we retrieved global colours and inferred stellar masses with a \citet[][]{Chabrier2003} initial mass function. A more detailed description of the corrections applied to this and other quantities, will be presented along their corresponding sections.

\section{Methods and Procedures}
\label{sec:method}

\subsection{Morphological Classification}
\label{sec:morpho-classification}

To carry out the morphological classification we retrieved images from the SDSS and DESI Legacy surveys, that were digitally post-processed to enhance internal structures (bars, spiral arms, rings, etc.), as well as external structures (low surface brightness spiral arms and tidal debris) in the $r-$band images. As a result, we generated image mosaics for each individual galaxy, containing colour composite images from SDSS and DESI\footnote{Note: in this paper we will refer as "DESI images", to those from the DESI Legacy Surveys}, and the DESI residual images after subtraction of a best surface brightness model. A more detailed description of our procedures and classification scheme are presented below.

\subsubsection{SDSS Image Preparation}

Our scheme for the visual classification of MaNGA galaxies includes a reprocessing of the SDSS legacy images. We retrieved the calibrated science frames from the SDSS DR12 survey \citep{Alam2015} and generated new $r-$band image mosaics by using the Montage software \citep{Jacob2010}. We improve on previous images by performing a new internal background matching and gradient removal to the frames before combining them into new mosaics. The final mosaics are centered on the galaxy coordinates, preserving the SDSS native pixel size of 0.396 arcsec/pix and extending farther out than the size of MaNGA galaxies. This processing is crucial for a correct performance of image enhancement algorithms for the visual identification of morphological structures. 
The depth of the images after the Montage processing is, in the worst case, the same as that of the archive SDSS images. However, depending on the number of images used to generate a new mosaic and on the signal-to-noise ratio, the depth may improve over that of the original SDSS images by factors up to half a magnitude in the $r-$band.

A following step was the generation of 'filter-enhanced' $r-$band  SDSS images for each MaNGA galaxy after testing Gaussian kernels of different sizes and orientations. Filter enhancing is a variant of the unsharp masking procedure that emphasizes high frequency spatial internal structure in the form of star-forming regions, bars, rings, and/or structure embedded into dusty regions. Additionally, the background was conveniently cancelled-out at the outskirts for identification of low surface brightness features. 

\subsubsection{DESI Legacy Surveys images}

Three wide-area optical imaging surveys were designed to provide targets for
DESI \citep{Dey2019}. The Dark Energy Camera (DECam; \citealt{Flaugher2015}) at the 4m Blanco telescope at the Cerro Tololo Inter-American Observatory obtained images with median FWHM seeing of a delivered image quality (DIQ) $\sim$ 1.3, 1.2, and 1.1 arcsec in the $g, r$, and $z$ bands, respectively, for the DECaLS survey. The Beijing-Arizona Sky Survey (BASS; \citealt{Zou2017}) obtained imaging with median DIQ of 1.6 and 1.5 in the $g$ and $r$ bands, respectively. Finally, the Mayall $z$-band Legacy Survey (MzLS) complemented the BASS $g-$ and $r-$band observations in the same sub-region of the Legacy Surveys with a median DIQ of $\sim$ 1 arcsec. The depth of the DESI imaging reaches $g = 24.0$, $r = 23.4$, and $z = 22.5$ AB mag for a point source target \citep{Dey2019}.   

The DESI quality requirements translate in principle into deeper images than the SDSS images, reaching a mean surface brightness limit of $\sim$26 for a 1$\sigma$ detection by using the corresponding $r-$band variance images (Cano-Diaz et al. in prep.),
more appropriate for the detection of internal and external morphological details. 

The $r-$band DESI Legacy images and the corresponding $grz$ composite colour images were retrieved for each MaNGA galaxy. The images were centered on the galaxy coordinates, preserving the DESI Legacy Surveys native pixel size and extending much farther out than the size of MaNGA galaxies to ensure the detection of low surface brightness features. Similar to the SDSS images, the $r-$band DESI images were also 'filter-enhanced' for visual identification of inner and outer morphological features.

In addition to the DESI legacy images, we also retrieved the residual images of the post-processing catalogue for the Legacy Surveys. This catalogue uses an approach to estimate source shapes and brightness properties. Each source was modeled by The Tractor package \citep[for more details, see][]{Dey2019} using a set of parametric light profiles: a $\delta$ function (for point sources); a de Vaucouleurs r$^{-1/4}$ law; an exponential disk; or a 'composite' de Vaucouleurs plus exponential. The best fit model was determined by convolving each model with the specific PSF for each individual exposure, fitting to each image, and minimizing the residuals for all images. The residual images provide valuable information for the identification of morphological features in the inner regions of galaxies. Note that we have retrieved images from the DESI Legacy Surveys Data Release 8, the available images at the beginning of this work.

\subsubsection {Morphological Considerations}
\label{sec:steps}

\begin{figure}
 \includegraphics[width=\columnwidth]{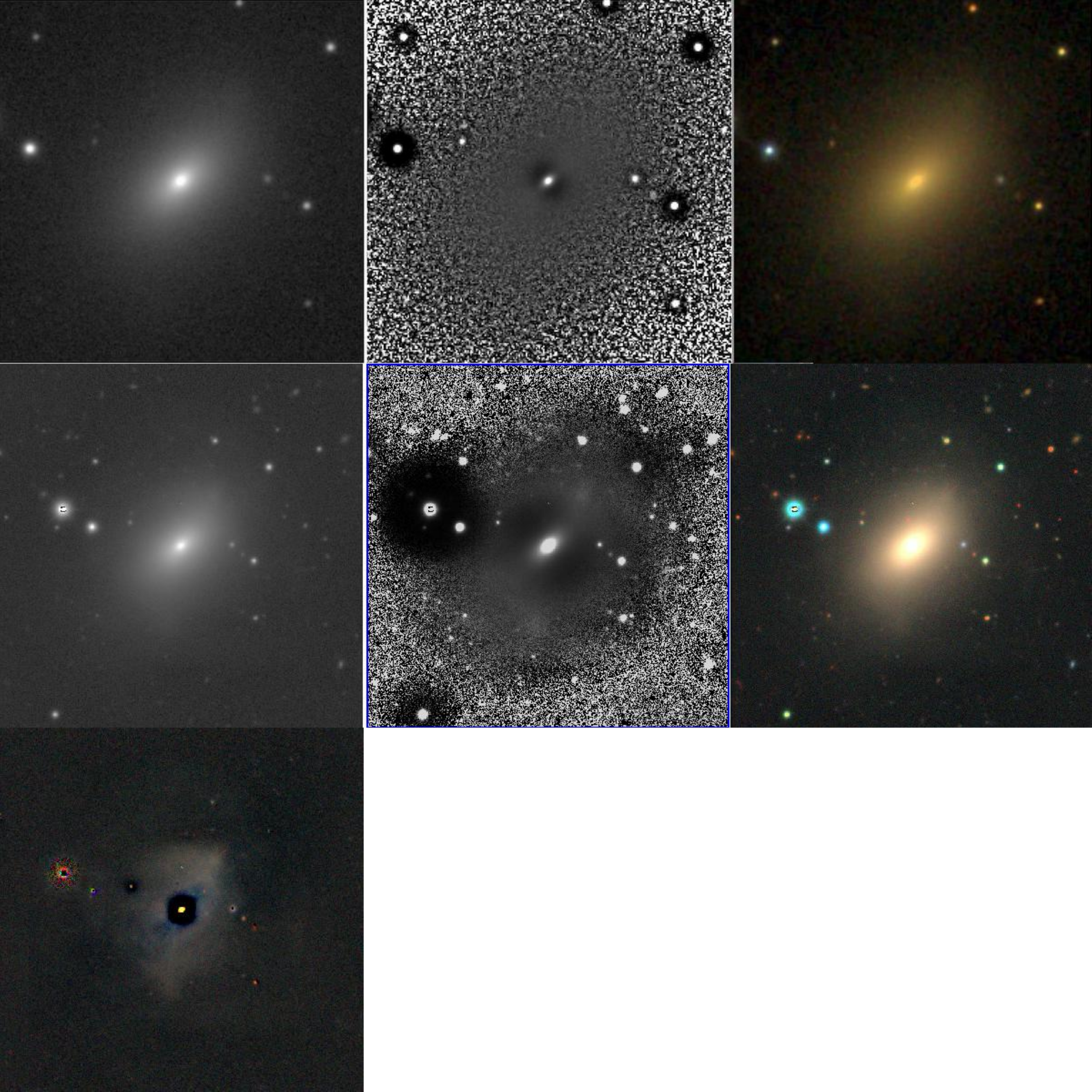}
 \caption{An example of our image processing of the  
 SDSS (upper panels) and DESI Legacy Surveys (middle panels) images, corresponding to the low-inclined galaxy manga-8728-12703. Each panel shows from left to right: a gray logarithmic-scaled $r-$band image, a filter-enhanced $r-$band image and the \emph{gri} or \emph{grz} composite colour images. {\it Lower panel:} Residual image (after subtracting the best fit surface brightness model) from the DESI post-processing catalogue. This galaxy was classified as S0.}
 \label{fig:mosaic}
\end{figure}

Our morphological classification followed the basic Hubble scheme (E, S0, Sa, Sb, Sc, Sd, Sm, Irr and their intermediate types), and it was implemented in a simple way. Figure~\ref{fig:mosaic} summarizes for an example galaxy (manga-8728-12703), the results of our digital processing to the SDSS and DESI images. The upper panel presents gray logarithmic-scaled $r-$band images (left), filter-enhanced $r-$band images (middle), and the corresponding $gri$ colour composite (right) SDSS images. Middle panels show similar results to the upper panels but for the DESI images. The lower panel shows a PSF-convolved residual image from the DESI post-processing catalogue. This image processing is useful for the identification of disk features, reinforcing the classification of manga-8728-12703 as S0.

Two of us (HHT and JAVM) carried out independent visual inspections to our set of mosaics for the DR15 MaNGA sample. The results were then compared between classifiers on a line by line basis and then homogenised. 
It should be mentioned that we tried the morphological evaluation of each galaxy when possible. This was not the case for the merging galaxies and strongly disturbed galaxies. However, there were also cases of relatively isolated galaxies, where the morphological evaluation was not clear. Although they were assigned a tentative morphological type, these galaxies were labelled 'unsure'. These are visually faint or diffuse objects, and some of them barely showing signs of an apparent disk (S-like) at different inclinations, and a few of low concentrations. We will discuss them in detail in Section~\ref{Sec:diffuse}. The most evident (bright) morphological tidal features were also identified. Again, the independent results were also compared and then homogenised.

Our filter-enhanced images in combination with the DESI residual images were considered as a useful tool for the identification of inner and outer morphological features. An internal weight (0/1) was given to the DESI residual images such that, they were considered for the classification (weight $=$ 1) only in cases where the features identified were similar between the residual and the filter enhanced images. This strategy allowed us for the recognition of early-type E galaxy candidates showing possible inner/intermediate size embedded disks, useful for separating E and S0 types as explained in the following section.
 
Finally, once all galaxies with possible bars were identified, one of us (MHE) carried out a new visual inspection for a further classification into bar families following the criteria of the Comprehensive de Vaucouleurs revised Hubble-Sandage (CVRHS) system \citep{Buta2015}. 
This is in contrast to other more simplified classification schemes which consider only the length and light prominence of the bar. The results on these bar families will be reported in a forthcoming paper of this series.  

\subsubsection {MaNGA E,S0,S0a galaxies}
\label{sec:ES0}

In \citet{HernandezToledo2008,HernandezToledo2010} early-type (E/S0/S0a) candidates were identified by combining a visual evaluation with a parametrization of their light distribution after an isophotal analysis. In the present paper we use that experience acquired and combine it with a purely light-based approach by \citet{Cheng2011}, where early-type galaxies are accommodated according to their light distribution into: bulges, smooth discs and non-smooth discs. Here we use this combined approach to assign galaxies into E, S0 and S0a Hubble types respectively.

\begin{figure}
 \includegraphics[width=\columnwidth]{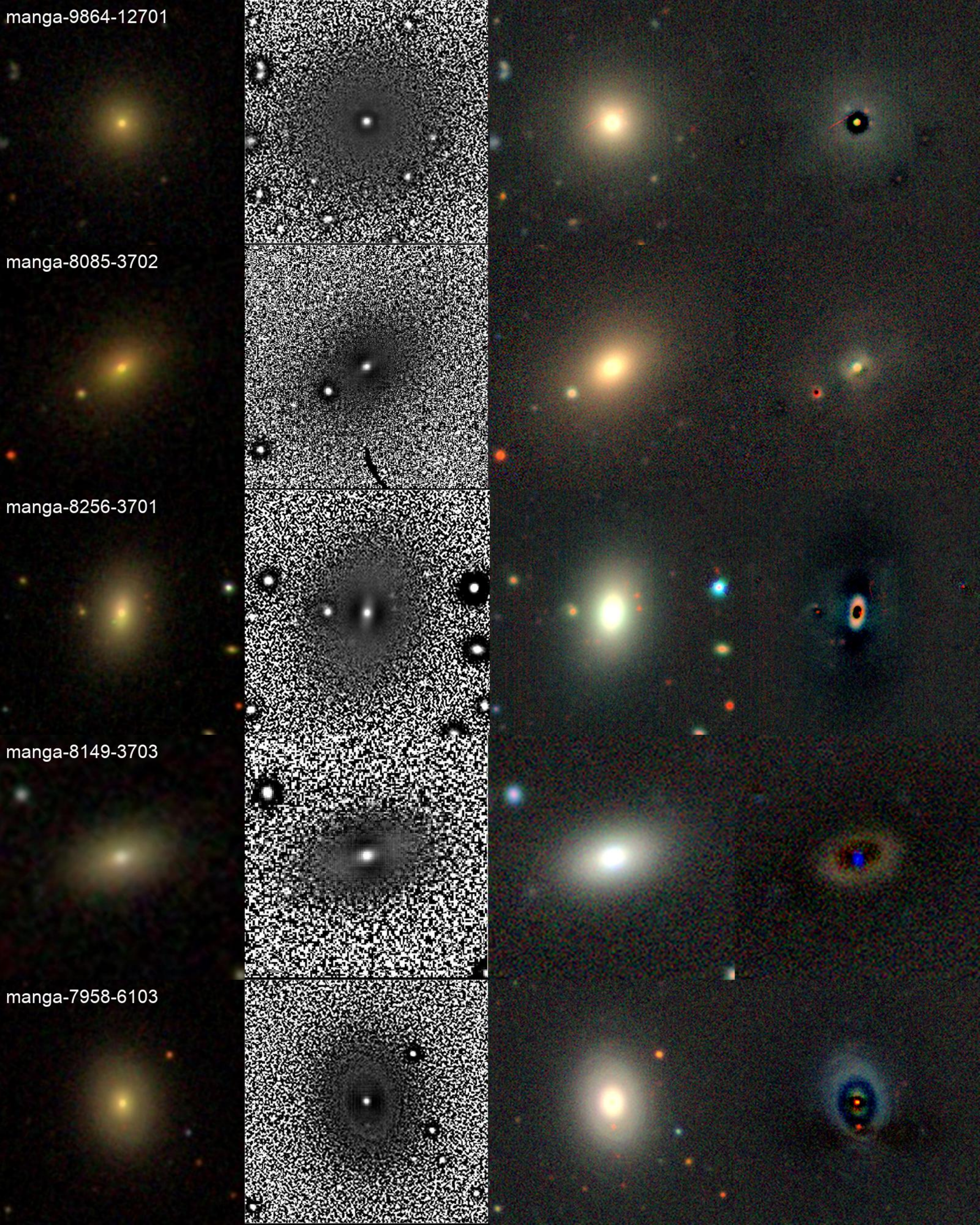}
 \caption{Mosaic examples showing our classification scheme for early type galaxies. From left to right, the \emph{gri} SDSS, the filtered-enhanced $r-$band, the \emph{grz} DESI, and the residual DESI images as described in Sec~\ref{sec:steps}. From top to bottom panels: 9864-12701 a pure elliptical (E) galaxy, 8085-3702 an elliptical with an inner disk (Edc), 8256-3701 a lenticular (S0) with an intermediate-size disk, 8149-3703 an S0 galaxy and 7958-6103 a transition S0a galaxy.}
 \label{fig:ES0S0a_mosaic2}
\end{figure}

Figure~\ref{fig:ES0S0a_mosaic2} illustrates the identification of early (E,S0,S0a) types. Each case is shown in an horizontal panel consisting of four images; a $gri$ SDSS image (left), a filter-enhanced $r-$band image (middle left), a $grz$ DESI image (middle right) and the corresponding residual DESI image (right) after subtracting a best fit model to its light distribution.

Elliptical galaxies were identified as those having high central concentration with a gradual fall-off in brightness at all radii and outer regions having no sharp edges (see upper-most panel of Fig~\ref{fig:ES0S0a_mosaic2}). However, in our experience Ellipticals are not composed of pure dominated bulges. With the help of our filter-enhanced images and the PSF-convolved DESI residual images, we were able to uncover Elliptical galaxy candidates showing the presence of possible small disks embedded in the inner region of the spheroid, dubbed as ellipticals with inner disks (Edc; second panel from top to bottom of Fig~\ref{fig:ES0S0a_mosaic2}).

We were also able to uncover early-type galaxies showing possible intermediate-size disks at different orientations embedded in an important fraction of the spheroid that can be either featureless or weakly featured (third panel from top to bottom of Fig~\ref{fig:ES0S0a_mosaic2}), dubbed as (S0) Lenticulars with intermediate-sized disks. More frequently, Lenticular galaxies were identified as those having a relatively prominent central light concentration with a sharp outer edge, where the light drops off drastically showing a relatively flat profile at intermediate radii (fourth panel from top to bottom of Fig~\ref{fig:ES0S0a_mosaic2}). There are cases 
of prominent outer rings evidencing the presence of an extended disk in these galaxies.

\citet{Graham2019} presented a modified galaxy classification scheme for local galaxies emphasizing the often overlooked continua of disc sizes in early-type galaxies. That schema encompasses nuclear discs in Elliptical (E) galaxies, intermediate-scale discs in Ellicular (ES) galaxies, and large-scale discs in lenticular (S0) galaxies. 

Although our image processing may be helpful to identify some edge-on inner or intermediate size disks, a more detailed image analysis must be considered in order to uncover disks in different orientations like in the E/ES/S0 schema by \citet{Graham2019}. Such an image processing effort is beyond the scope of the present classification. A benefit of using the current IFU MaNGA data is that such disks could be readily seen after a kinematic analysis, unless they are almost perfectly face-on \citep[c.f.][]{Emsellem2011,Cappellari+2011,Krajnovic2013}

Figure~\ref{fig:ES0S0a_mosaic2} also illustrates the ability of our processed images to uncover the outer regions of disks and their associated spiral-like features in transition S0a galaxies (lower panel).  

A limitation of our visual procedures for separating between E and S0 types is that our classification still allows for a possible contamination of the E sample by face-on S0s. In our experience, a more quantitative approach as shown by \citet{HernandezToledo2008,HernandezToledo2010}, in combination with a detailed 2D parametric decomposition of the DESI images, could be more appropriate to identify face-on S0 galaxies among E galaxies. At present, following the classification here described, only the most evident S0 cases were separated from our E sample.

\subsubsection {MaNGA Sa-Irr galaxies}
\label{sec:late-gal}

\begin{figure}
 \includegraphics[width=\columnwidth]{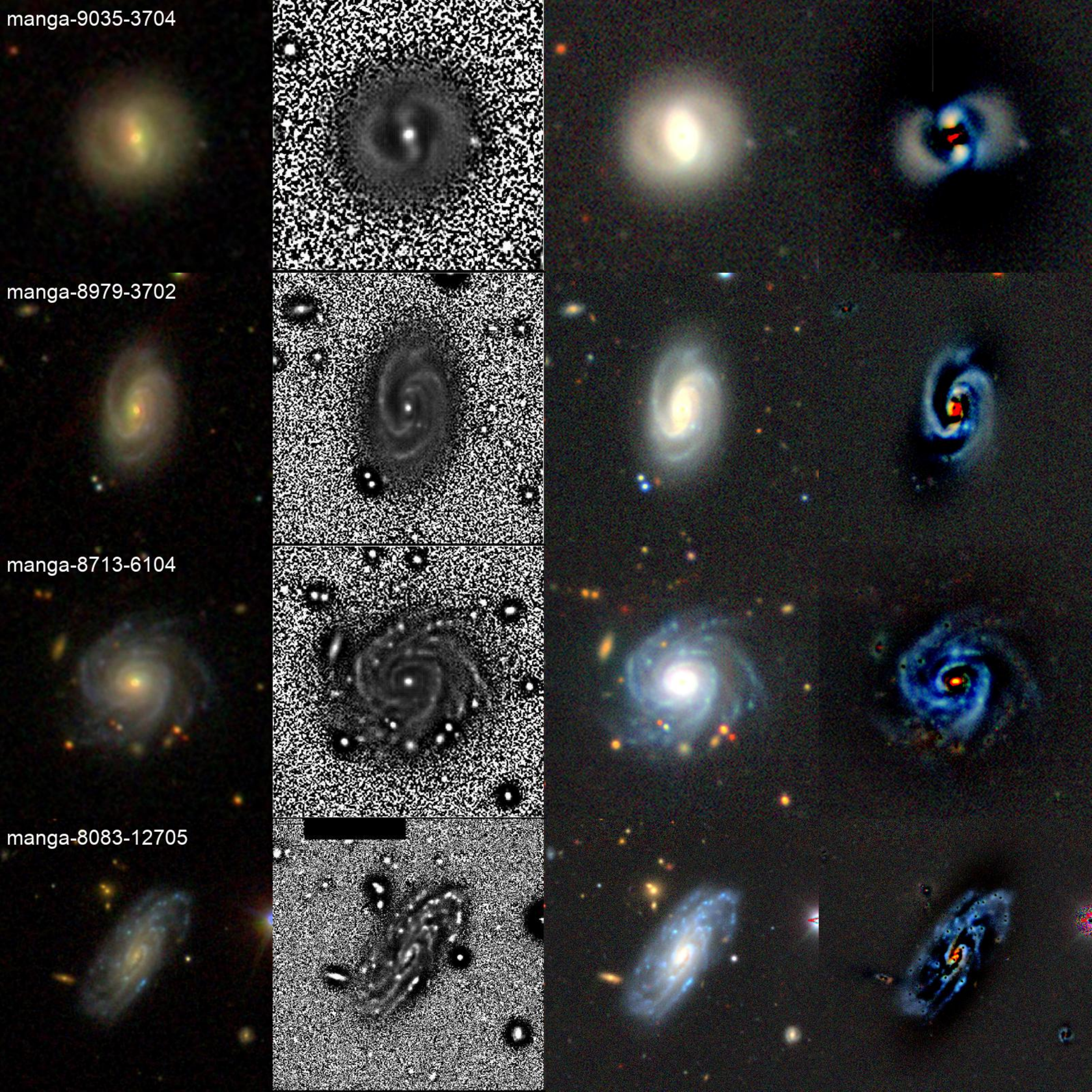}
 \caption{Mosaic examples illustrating our classification scheme for late type galaxies. From top to bottom panels: 9035-3704 (Sa), 8979-3702 (Sb), 8713-6104 (Sc), and bottom panel 8083-12705 (Sd). Columns are similar to Fig~\ref{fig:ES0S0a_mosaic2}}
 \label{fig:S_mosaic2}
\end{figure}

Figure~\ref{fig:S_mosaic2} shows our combination SDSS and DESI processed images for the identification of morphological features in low inclination Sa- to Sd- types, similar to Figure~\ref{fig:ES0S0a_mosaic2}.   

Low inclination Sa-Sd types (from top to bottom panels) were identified as those showing gradually less dominant central light concentration and the presence of a definite disk containing weak to strong disk signatures. We figured out the degree of resolution of the spiral arms into fragments and how tightly or loosely wound they were when possible. The filter enhanced (second column) and  the residual images (fourth column) were useful to that purpose. At the end of the late-type spiral sequence we also identified faint central bulges and the flocculency of the arms made up of individual stellar clusters and clumps typical of Sd (SBd) types (bottom panel), up to galaxies with practically no bulge component from irregular appearance Sm (SBm) to highly irregular shapes Im types. 

\subsubsection {Highly inclined galaxies}

The advent of recent IFU surveys \citep[e.g. CALIFA][]{Sanchez2012} have shown the potential that highly inclined galaxies offer for the analysis of properties in disk galaxies including their structure, kinematics, and stellar content \citep[e.g.,][]{Levy2019, Jones2017, Bizyaev2017, Ho2016, LopezCoba2019}.
In our morphological evaluation we visually confirmed lenticular and spiral galaxies with inclinations $>$70$^{\circ}$ using a-priori known semi-axis $b/a$ ratio taken from the NSA Atlas. At this inclination limit, the sample includes not only nearly perfect edge-on galaxies, but also moderately inclined galaxies, where various morphological features may or not be distinguishable, however the bulge region can be detected against the strong contrast of the background allowing us for a visual estimate of the bulge/disk ratio and thus for a rough visual morphological classification. 

\begin{figure}
\includegraphics[width=\columnwidth]{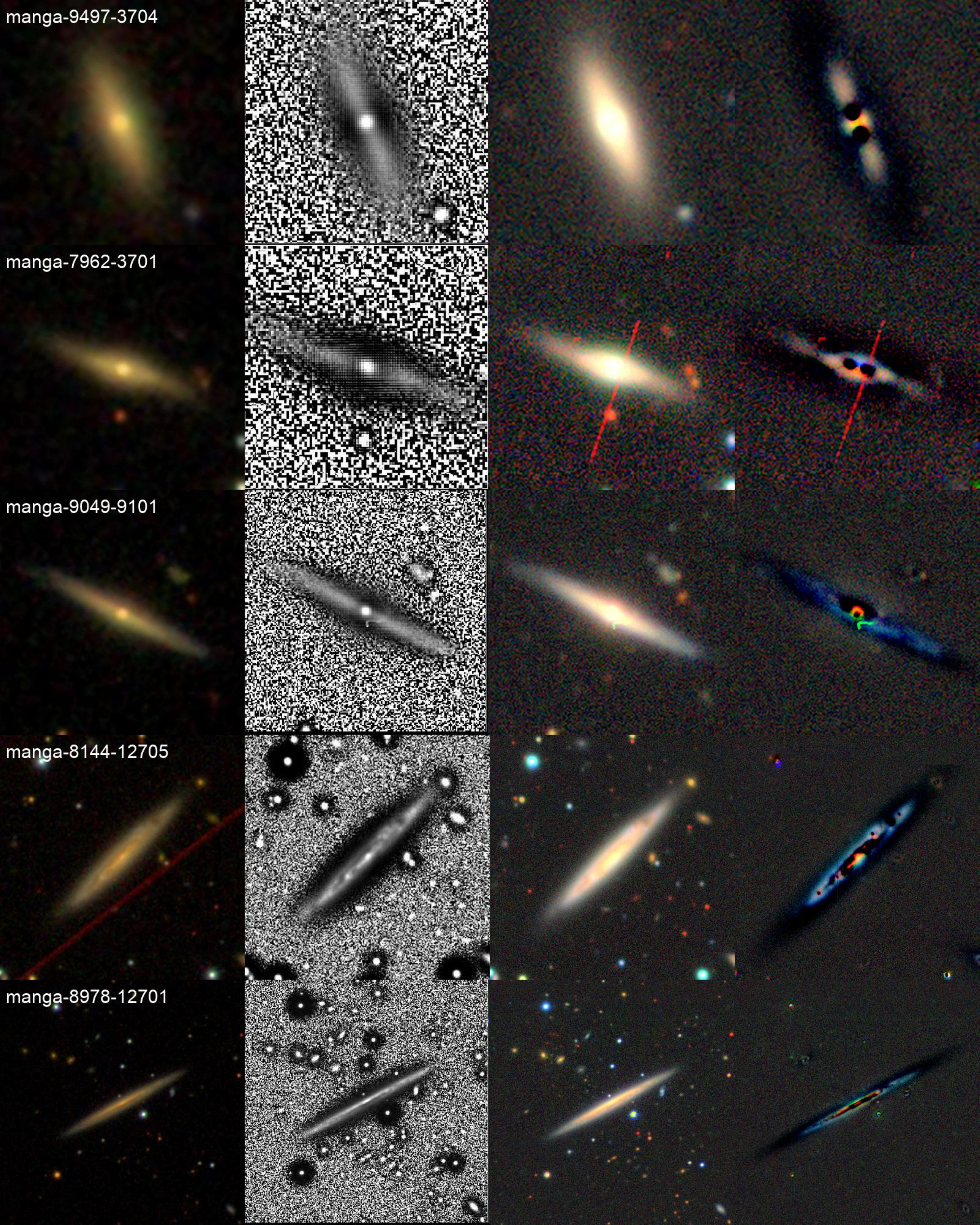}
 \caption{Mosaic examples to illustrate our classification scheme of nearly edge-on galaxies (i $>$70$^{\circ}$). From top to bottom panels: 9497-3704 (S0), 7962-3701 (Sa),  9049-9101 (Sb), 8144-12705 (Sc), and 8978-12701 (Sd). Note the {\bf absence} of a visible bulge and internal structure evidence to later types. Columns are similar to Fig~\ref{fig:ES0S0a_mosaic2}.}
 \label{fig:Incl-mosaic}
\end{figure}

Figure~\ref{fig:Incl-mosaic} is a four-column mosaic showing some examples of edge-on and almost edge-on S0- to Sd- types. This figure illustrates the advantages of our filter enhanced DESI images (second column) and residual DESI images (fourth column) to identify morphological features. Edge-on S0 galaxies are identified as those having a prominent to moderately prominent bulge central region, no large-scale clumpy structure along the disk, and no signs of spiral-like features emerging from the edges (upper-most panel). Likewise, edge-on early-type spirals (two intermediate panels) and late-type spirals (two lower panels) can be identified according to the apparent bulge/disk estimates and the presence of features along and at the extremes of the disk.

\subsubsection{Bar Classification}
\label{Sec:BarClasificaion}

An important task in the morphological evaluation of MaNGA galaxies is the identification and estimate of the bar fraction. Bars play an important role in the dynamics of disks and the IFU observations from the MaNGA survey offer an unique opportunity to test this role and their relation to other properties of galaxies in more detail.

Barred galaxy candidates in the MaNGA sample were identified as part of the outlined procedures. In practice a first assignment considered no bars, suspected bars and definite bars along the different Hubble types (for S0 and spirals). Non-barred spiral galaxies (SA) show no evidence of a bar in general, although high inclination can cause  miss-identifications. However, even in these cases, it is still possible to know whether there is a (rather strong) bar by finding evidence of a boxy/peanut/X-shaped bulge, which has vertical structure extending above and below the plane of the disks \citep[see for example][]{Buta2015, Laurikainen2014,Athanassoula2015,HerreraEndoqui2017}. Thus, in cases of edge-on galaxies where boxy/peanut/X-shaped bulge are identified, the galaxy is classified as strongly barred (B).  

Notice that the identification of bars presented here is mainly based on the most evident and easily identifiable cases and omits other possibilities like nuclear bars, out of our resolution capacity and small bars aligned with spiral arms as illustrated in more detailed studies based on isophotal and Fourier analysis \citep[e.g.][]{DiazGarcia2016}

\subsubsection{Tidal Features Identification}
\label{Sec:Tides}

The present and next generation of deep extragalactic surveys (DESI; Euclid, \citealt{Laureijs2011}; HSC-SSP, \citealt{Aihara2018}; LSST, \citealt{Ivezic2019},  and others) are stimulating new strategies for the detection of stellar structures such as streams, stellar shells, rings, plumes and other tidal features associated to galaxies. The origin of these structures is also the subject of a detailed analysis from galaxy simulations \citep[e.g. ][and references therein]{RodriguezGomez2016, Mancillas2019} since their properties reveal important information on the dynamics and assembly history of galaxies \citep[e.g. ][]{Johnston2008, Belokurov2017}. Furthermore, recent IFU based studies like those provided by the MaNGA Survey or the MUSE instrument at the VLT \citep[e.g.][]{Bacon2015} are complementing this findings, providing key information on the kinematic and chemical properties of these stellar structures \citep[e.g. ][]{Thorp2019, Fensch2020}.

\begin{figure}
 \includegraphics[width=\columnwidth]{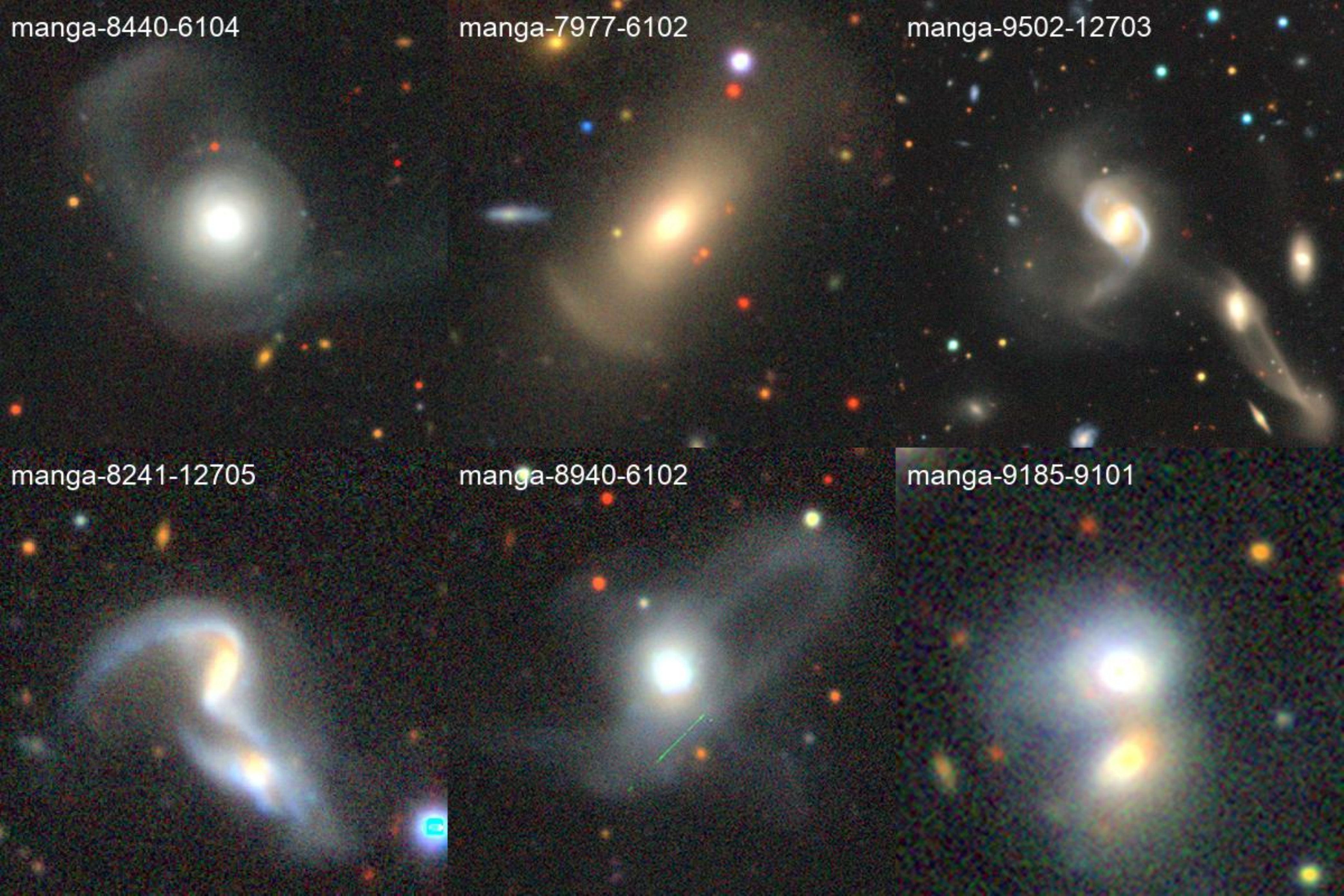}
 \caption{Examples of galaxies identified with different bright tidal features using the \emph{grz} DESI images. Manga-8440-6104 (E type) shows filaments and streams. Manga-7977-6102 (S0 type) presents shells and fans. Bridges and tails are observed in Manga-9502-12703 (SBb type) and Manga-8241-12705 (SBc type). Streams and filaments are presented in Manga-8940-6102 (S0 type). Finally Manga-9185-9101 (S0 type) shows evidence of broad fans. Note that {\it i)} tidal identification was carried out considering only the DESI colour-composite image, and {\it ii)} in cases of very close pairs, the classification was assigned to the galaxy closer to the retrieved MaNGA position.}
 \label{fig:tides_example}
\end{figure}

An important goal in our morphological characterization of the MaNGA DR15 sample was the visual identification of the most evident (bright) tidal features. To this purpose only the original colour composite $grz$ DESI images were considered. Figure~\ref{fig:tides_example} shows a variety of tidal features, to mention, streams, filaments, shells, fans, and structures probably produced by close or nearby interactions either from the primary galaxy as well as those from the companion(s) being stripped. Notice however that we are only reporting the existence or not of such structures in a binary flag (\emph{Tides} $=$ 1, 0) with no explicit reference to any of the above mentioned classes. A further analysis through a more appropriate image processing, making an explicit mention to any of tidal classes and exploring their origin or nature will be reported elsewhere.

\subsection{Volume Corrections and Completeness}
\label{sec:Vcorr}
 
A fundamental parameter behind the selection of MaNGA galaxies is the stellar mass, such that a roughly flat distribution in $\log M_{\ast}$ was imposed \citep[][]{Wake2017}. 
The MaNGA selection also makes this sample incomplete in colour and morphology. 
If we were to perform a statistical characterization of the MaNGA sample in terms of stellar mass, a way to proceed would be to compare this sample to other volume-limited sample that is complete in stellar mass. Based on that comparison, we introduce for the MaNGA sample completeness correction factors $f_m$ (defined as the expected fraction of galaxies as a function of stellar mass, colour and redshift for the MaNGA sample) to recover the Galaxy Stellar Mass Function (hereafter GSMF).   

Next, we briefly describe the volume and completeness correction factors $f_m$ for the MaNGA sample. 
As shown in \citep[][see also \citealp{Rodriguez-Puebla+2020}]{Sanchez+2019}, the volume
correction for the $j$th galaxy in the MaNGA sample is given by
\begin{equation}
   \frac{1}{V_j}=  \frac{w_i(M_\ast,g-r, z)}{f_m(M_\ast,g-r, z)dV (z)},
\end{equation}
where $w_i(M_\ast,g-r, z) = N_{i}/\sum_iN_i$; here $N_i$ is the number of 
galaxies in the rectangular cuboid with sides
$\log M_{\ast}\pm d\log M_{\ast}/2$, $(g-r)\pm d(g-r)/2$ and $z\pm dz/2$ and the 
summation over $i$ refers to redshift. The correction factors $f_m$ are given by
\begin{equation}
    f_m^{-1}(M_\ast,g-r, z) =\frac{1}{\phi(M_{\ast},g-r)} \sum_{i}\frac{w_iN_{i}}{d\log M_{\ast}dV(z_i)},
    \label{eq:weight}
\end{equation}
where $\phi(M_{\ast},g-r)$ is the observed GSMF as a function of colour $g-r$ 
based on the $NSA$ sample \citep{Blanton2017}.

We calculate volume and completeness correction factors for every galaxy in the MaNGA sample 
by creating a grid of 30 redshift bins within $0.01<z<0.15$ and 50 stellar mass bins 
within $10^8 < M_\ast/M_\odot < 10^{12}$ and 5 colours bins within $0<g-r<1.2$. The details and methodology will be presented in a forthcoming paper (Calette et al. in prep, see also \citealp{Sanchez+2019} and \citealp{Rodriguez-Puebla+2020}).

The MaNGA survey also included various ancillary programs. The current DR15 sample includes data from 16 ancillary programs. We have identified 182 galaxies from such programs in our sample that have been removed due to possible bias in the weights; however, we inspected the differences between considering or not ancillary galaxies are practically indistinguishable. Due to the nature of the sample, these corrections are trustworthy for galaxies with $\log(M_\ast/M_\odot)>$ 9 omitting galaxies with masses below this value. Therefore in the next sections when we refer to volume-corrected fractions, the number of galaxies involved accounts to 4411 galaxies.

\begin{figure*}
 \includegraphics[width=0.8\textwidth]{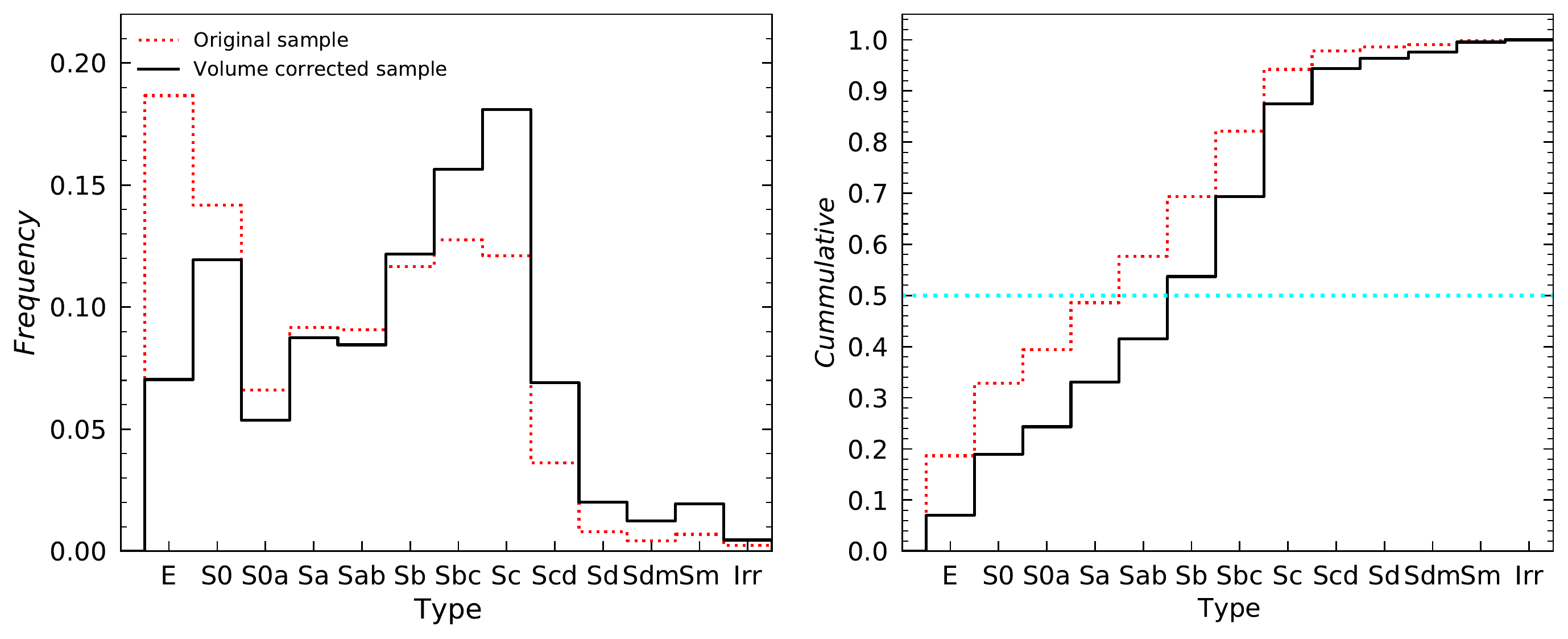}
 \caption{{\it Left panel:} Frequency distribution of morphology (normalized to the total number of galaxies) for the original (dotted-red) and volume-corrected (solid-black) MaNGA DR15 sample. {\it Right panel:} The corresponding cumulative distributions. The horizontal dotted line (cyan) indicates the cumulative 50\% of the sample. NOTE: For the volume corrected histogram, only galaxies with $\log (\ms/M_{\sun})>$9 are considered and ancillary galaxies were removed.
 }
 \label{fig:hist_morpho}
\end{figure*}

\section{Results}
\label{sec:results}
 
\subsection{Morphological Content}

From 4696 galaxies originally considered in the MaNGA DR15 sample, we identified 4619 having individual non-repeated IDs, aside of 5 stars, yielding a final sample of 4614 galaxies (and 4411 when volume corrections are considered). Galaxies immersed within scattered light halos or diffraction spikes as well as those showing light contamination coming from other large nearby galaxies were also given a tentative classification when possible. Table~\ref{tab:morpho} reports the morphological evaluation.

Column 1 and 2 state the assigned types with their standard code number. Column 3 reports the observed number of galaxies corresponding to each morphological type , while Column 4 reports the fraction with respect to the total sample for a given type. 
Column 5 reports the volume-corrected fractions for the MaNGA DR15 galaxies using the weights given by Eq. (\ref{eq:weight}). Finally, the lower panel of Table~\ref{tab:morpho} reports the fractions of: Edge-on  galaxies ($> 70^{\circ}$), galaxies with visible tidal features, barred galaxies, and Faint-Diffuse-Compact galaxies dubbed as 'unsure' (see Section~\ref{Sec:diffuse}).  

\begin{table}
  \begin{center}
    \caption{The morphological evaluation of the MaNGA sample. Number and fraction with respect to the total sample is presented, as well as the inferred fraction when volume-corrections are applied. Notice that {\it i)} T-Type NC (no classification) corresponds to strongly perturbed galaxies in interacting systems, where a T-Type is not possible to assign, {\it ii)} volume-corrected fractions are estimated only for galaxies with $M_{\ast}>10^{9}M_{\odot}$, where ancillary galaxies were omitted.}
    \label{tab:morpho}
    \begin{tabular}{c c c c c}
    \hline
      \textbf{T-Type} & \textbf{T-Type} & \textbf{N} & \textbf{Fraction} & \textbf{Volume-corr} \\
             & \textbf{code}\tablefootnote{NOTE: in order to make plots and calculations in a continuous range, for this paper T-Type codes -5 and -2 are replaced by -2 and -1 respectively.} 
             &   &  & \textbf{fraction} \\
      \hline
E & -5 & 860 & 0.186 & 0.07 \\
S0 & -2 & 653 & 0.142 & 0.119 \\
S0a & 0 & 304 & 0.066 & 0.054 \\
Sa & 1 & 422 & 0.091 & 0.087 \\
Sab & 2 & 418 & 0.091 & 0.084 \\
Sb & 3 & 537 & 0.116 & 0.121 \\
Sbc & 4 & 588 & 0.127 & 0.156 \\
Sc & 5 & 558 & 0.121 & 0.181 \\
Scd & 6 & 167 & 0.036 & 0.069 \\
Sd & 7 & 37 & 0.008 & 0.02 \\
Sdm & 8 & 20 & 0.004 & 0.012 \\
Sm & 9 & 32 & 0.007 & 0.019 \\
Irr & 10 & 11 & 0.002 & 0.005 \\
NC & 11 & 7 & 0.002 & 0.002 \\
    \hline
    Edge-on ($> 70^{\circ}$) & & 776 & 0.2 & 0.24\\
    Tides & & 751 & 0.16 & 0.14\\
    Barred & & 1727 & 0.46 &  0.47\\
    Faint-Diffuse-Compact & & 387 & 0.08 & 0.14\\
    \hline
    \end{tabular}
  \end{center}
\end{table}

Figure~\ref{fig:hist_morpho} summarizes the results from Table~\ref{tab:morpho}. The left panel shows the frequency distribution (normalized to the total) of \type\ for the original (red-dotted line) and volume corrected (black-solid line) samples. Notice that the histogram of the original sample shows dominance of early types. However, given the dependence of the volume-correction factors $f_m$(\ms,$g-r$, $z$) on \ms, colour, and $z$, the corrected distribution shows a major contribution coming from intermediate-to-late spiral types, which are typically of lower masses and bluer colours than early-type galaxies. This illustrates the correction applied to the MaNGA DR15 sample to be consistent with the local GSMF.

The right panel shows the corresponding cumulative distributions, also normalised to the total number of galaxies, before (red-dotted line) and after corrections for mass completeness in the sample volume (black-solid line). The horizontal cyan dashed line indicates the 50\% cumulative fraction of the sample, reached about Sb types.

Left panel of Figure~\ref{fig:hist_morpho} evidences a bi-modal distribution for the morphological types. Considering volume-corrected frequencies, approximately one mode corresponds to E--S0 types (18.9\%) and the other mode corresponds to Sab--Irr types (66.7\%), with a peak in the Sc types. The minimum in the distribution corresponds approximately to S0a--Sa types (14.1\%).
According to these results, we can nominally separate the galaxy population into three groups: early-type (E--S0), intermediate-type (S0a--Sa), and late-type (Sab--Irr) galaxies, ETG, ITG, and LTG, respectively.
Notice that the bulk of the morphological mix (94.1\%) is enclosed in the E--Scd types with a sudden decrease (5.6\% ) from Sd types and later. 
It is well known that Sd and later type galaxies are mostly low-mass or dwarf galaxies. Therefore, their abundance is expected to increase at masses lower than $\sim10^9$ \msun, which are below the limit of the current sample.

\subsection{The joint distribution of Hubble Type and stellar mass}
\label{sec:mass_Hubble}

Table~\ref{tab:mass} summarises the global relation between stellar mass and morphological types \type\ for the MaNGA DR15 sample. The stellar masses are adopted from the NASA-Sloan Atlas (NSA) catalogue \citep[][S\'ersic fluxes]{Blanton2011}, using $h=0.71$ Hubble constant. In the upper panel (original sample), Column 1 shows the stellar mass in intervals of 0.5 dex bins from $\log (\ms/M_{\sun})$ = 9 to $\log (\ms/M_{\sun})$ = 12. Columns 2 to 14 present, for a given mass interval, the numbers and fractions (in parentheses) corresponding to each morphological type \type\ from E to Irr. The lower sub-panel provides for each column, the number and fraction of galaxies per type \type, and the corresponding average and sigma values of $\log (\ms/M_{\sun})$. The bottom panel shows the corresponding fractions, similar to the upper panel, but after applying volume corrections.

\begin{figure*}
 \includegraphics[width=0.8\textwidth]{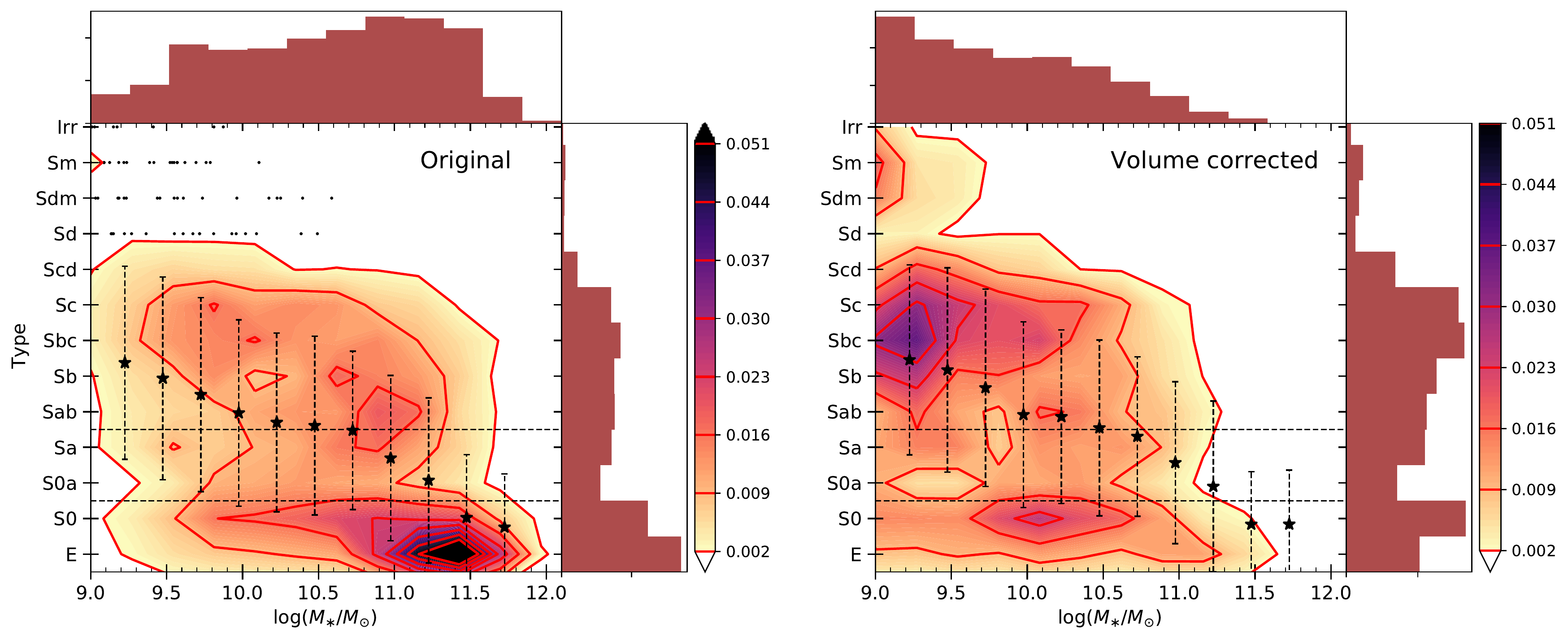}
 \caption{Bi-variate distribution of the MaNGA galaxies with $i\le70^\circ$ in the \type\ versus \ms\ diagram, not taking (\textit{left panels}) and taking (\textit{right panels}) into account volume completeness corrections. The bi-variate densities are shown with the isocontours and colours in linear scale (see colour palette for the normalized density). The top and right histograms in each panel show the marginalized 1D distributions. Notice how the stellar mass distribution changes dramatically after applying the weights for volume correction. However, the means and standards deviations (filled stars and dotted error bars) remain similar for both cases. Horizontal dashed lines show the transition zone of the bi-modal distribution.}
 \label{fig:mass_morpho}
\end{figure*}

Table~\ref{tab:mass} indicates some dependence of the morphological type on \ms. Most massive galaxies are mostly of early  types, while the less massive ones are commonly late-type spirals. The most massive galaxies, $>2\times 10^{10}$ \msun, are on average of early type (E--S0); the least massive ones,  $<5\times 10^{9}$ \msun, are on average of very late types (Sbc--Irr), while the galaxies of types S0--Sb are dispersed in a large range of masses.

Table~\ref{tab:mass} reveals that most of the MaNGA DR15 galaxies have been selected within $\pm$ 1.5 orders of magnitude in stellar mass around $\log (\ms/M_{\sun})$ $\sim$ 10.0 with an abrupt decrease in the number of galaxies at the low mass end. The $i$-band absolute magnitude selection in the MaNGA sample drives to a morphological mix increased in early-type galaxies and decreased in late-type ones as seen in Figure~\ref{fig:hist_morpho} (blue dotted histograms) and reported in Table~\ref{tab:morpho}. 
Also, Table~\ref{tab:mass} highlights a poor representation of dwarf galaxies, $\log (\ms/M_{\sun})$ < 9, in the MaNGA sample.\footnote{An approved MaNGA ancillary project for dwarfs was proposed by Cano-Diaz to complement the actual sample and it is being analyzed as part of the MaNDALA survey (Cano-Diaz et al., in prep.)}
Note that, since our volume completeness analysis was carried out by considering galaxies of masses larger than $\log (\ms/M_{\sun})\approx$  9, in the lower panel of Table~\ref{tab:mass} we can not estimate the expected fractions for galaxies of masses below $\log (M_{\ast}/M_{\sun})=$ 9. Finally, whether additional properties, such as those associated with the environment, are contributing or not to the observed morphological mix and their fractions in the MaNGA sample, is beyond the scope of the present paper and will be addressed in a forthcoming paper in this series.

Figure~\ref{fig:mass_morpho} shows the bi-variate distribution of \ms\ and \ \type\ with iso-density contours and colours {\it linearly spaced}; their labels are indicated in the colorbar. In the left-hand panel, the distribution is for the original sample, while in the right-hand panel, for the sample corrected to be complete in volume above $\log(\ms/\msun)\approx 9$. To avoid large uncertainties from very inclined galaxies, we exclude from both panels galaxies more inclined than $70^\circ$. Note that from the darkest to lightest colours there is a difference of a factor of more than 20.

The effect of the volume correction in the mass distribution is evident from the upper mass histograms in both panels. When comparing the left and right panels, it is clear that the number densities in the joint distributions change significantly after applying the weights to retrieve a sample complete in volume. However, the averages and standard deviations in stellar mass bins for all galaxies (stars with error bars) are roughly the same in the left and right panels since the weights at a given \ms\ are roughly the same, although we remember that our weights depend also on the ($g-r$) colour and $z$.

Once corrected by volume completeness, the right-hand panel of Figure \ref{fig:mass_morpho} shows a bi-modal joint distribution in the \type--\ms\ diagram, with a broad cloud of early and late spirals, Sab--Sd, highly populating the low- and intermediate mass range, $\log(\ms/\msun)\lesssim 10.3$, and a sequence of early-type galaxies (E--S0) populating a broad mass range, but dominated by massive galaxies, $\log(\ms/\msun)\gtrsim 10$. The abundance of very late-type galaxies (Sdm--Irr) is scarce and limited to low masses ($\log(\ms/\msun)\lesssim 9.7$); for masses lower than the surveyed by MaNGA ($<10^{10}$ \msun), the abundance of these galaxies is expected to increase dramatically.  The transition zone between the two modes mentioned above is around the S0a--Sa types (see the histogram on the right side of the panel) and it is nearly independent of mass. Thus, the  \type--\ms\ diagram can be nominally divided into the cloud of LTGs, the sequence of ETGs and the mass-independent valley of ITGs. The horizontal dotted lines in Figure \ref{fig:mass_morpho} indicate the region corresponding to this valley.

On average, the lower the mass, the later is the morphological type but with a large dispersion, see the averages (stars) and standard deviations in stellar mass bins. At the high-mass end, $\log(\ms/\msun)>11$, the majority of galaxies are ETGs (E--S0). At the low-mass end, there can be galaxies of all morphologies, but dominate by far the LTGs. 
The bi-modal distribution of galaxies seen in the \type--\ms\ diagram resembles roughly the well-known bi-modality in the colour--\ms\ diagram \citep[e.g.,][]{Weinmann+2006,Blanton2009}. In \S\S\ \ref{sec:color-age} we will discuss some details and the implications of this result, including also the light-weighted ages calculated from the spectral information of MaNGA galaxies.

\subsection{Bars: Morphology, Mass and Colour of their Host Galaxies}
\label{sec:bars}

Our results indicate that out of a sample of 3755 disk galaxies in the MaNGA DR15 sample, 1727 galaxies show evidence of a bar, yielding to a visual bar fraction of 46\%, changing slightly to 46.8\% when considering volume-corrected fractions. In order to compare with other results in the literature, we split into (S0-Sb) and (Sbc-Irr) disks. These fractions are 57.3\%(989) and 42.7\%(738) respectively, while the corresponding volume-corrected fractions are 45\% and 55\%. 
 
Our bar fraction is similar to that reported by \citet{Barazza2008} (48-52\%) and \citet{Aguerri2009} (45\%), who analysed optical SDSS images in local samples and used quantitative methods (ellipse fitting and Fourier analysis). When these studies are extended to the Near-InfraRed, these estimates can reach up to about two-thirds of nearby galaxies \citep{Eskridge2000, DiazGarcia2016}, or even higher \citep[][80\%]{Buta2015}.

Table~\ref{tab:bars} summarises the results of our visual inspection for bars in terms of morphological types. Column (1) presents the morphological T-types. Column (2) is the total number of barred galaxies and Column (3) presents, for each morphological bin, the number of barred galaxies normalized to the total number of bars in the MaNGA sample, before volume-correction and after volume corrections (right-after in parenthesis). At the bottom panel, additional to ETG (only S0 in this case), galaxies were accommodated into two morphological groups emphasizing the fractions of bars at ITG and LTG. 
Table~\ref{tab:bars} shows that bars are present in all Hubble types with the vast majority 96.6\% (93\%) hosted in S0--Scd types and only a very small fraction 3.4\% (7\%) in very late types (Sd--Sm).  
Notice, the volume-corrected fractions in parentheses.

\begin{table}
  \begin{center}
    \caption{Number of barred galaxies according to their morphology, following by the fraction to the total barred galaxies (and volume-corrected fraction in parentheses). Bottom panel additionally shows the fractions for the ITG and LTG groups.}
    \label{tab:bars}
    \begin{tabular}{c|c|c}
      \hline
      \textbf{Type} & \textbf{N$_\emph{Bar}$} & \textbf{Fraction}   \\
      \hline
S0 & 157 & 0.09  ( 0.07 ) \\
S0a & 107 & 0.06  ( 0.05 ) \\
Sa & 176 & 0.1  ( 0.08 ) \\
Sab & 259 & 0.15  ( 0.11 ) \\
Sb & 290 & 0.17  ( 0.13 ) \\
Sbc & 313 & 0.18  ( 0.19 ) \\
Sc & 269 & 0.16  ( 0.2 ) \\
Scd & 98 & 0.06  ( 0.09 ) \\
Sd & 19 & 0.01  ( 0.02 ) \\
Sdm & 16 & 0.01  ( 0.02 ) \\
Sm & 21 & 0.01  ( 0.02 ) \\
Irr & 2 & 0.0  ( 0.0 ) \\
\hline
Total & 1727 & 1.0  ( 1.0 ) \\
\hline
S0a-Sa & 283 & 0.16  ( 0.13 ) \\
Sab-Irr & 1287 & 0.75  ( 0.8 ) \\
      \hline
    \end{tabular}
  \end{center}
\end{table}

\begin{figure*}
 \includegraphics[width=0.8\textwidth]{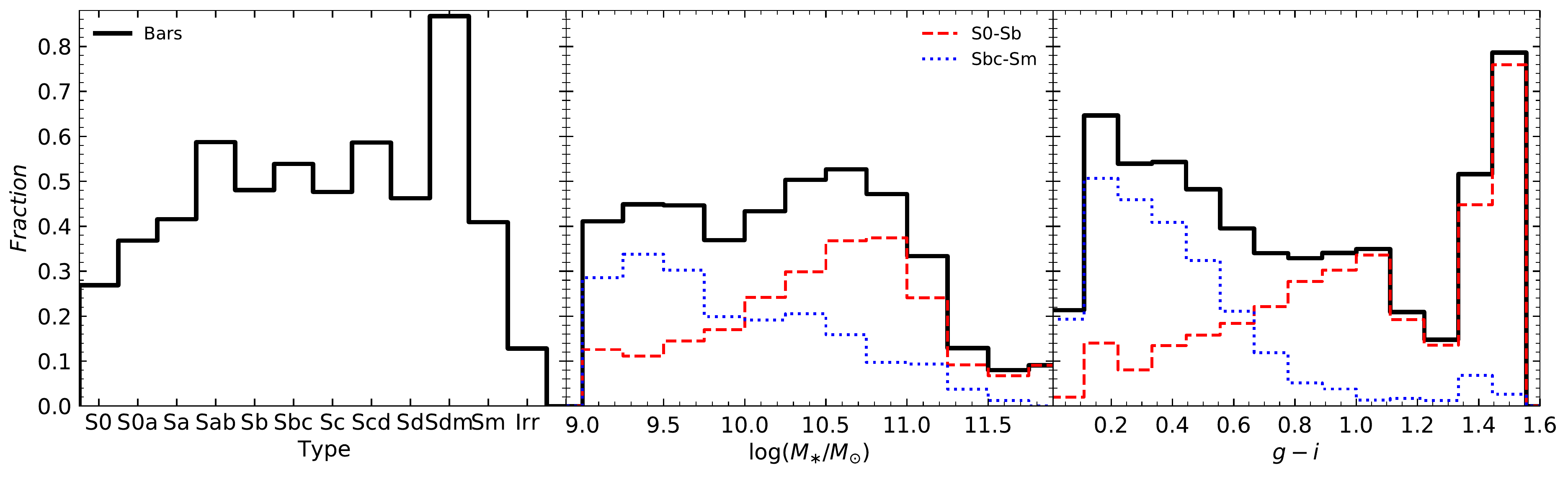}
 \caption{Volume-corrected fractions of barred galaxies (solid-black) as a function of morphology ({\it left}), stellar mass ({\it middle}) and \emph{g--i} colour ({\it right}). These fractions are also split into S0--Sb (red-dashed line) and Sbc--Sm types (blue-dotted line) types. Note: fractions are defined as the number of barred galaxies in each bin divided by the total number of galaxies in that bin.}
 \label{fig:Bar_fam}
\end{figure*}

Figure~\ref{fig:Bar_fam} (left panel) shows the volume-corrected fraction of barred galaxies as a function of morphological types. Similarly, the middle and right panels show the fractions in terms of stellar mass and $g-i$ colour. These panels emphasize the morphological contribution coming from the (S0--Sb) and (Sbc--Sm) morphological groups (red dashed and blue dotted lines, respectively) in order to compare with \citet{Nair2010b}. Colour corrections are described in Sec~\ref{sec:color-age}.

The left panel of Figure~\ref{fig:Bar_fam} shows that the barred fraction fluctuates around a level of 50\% from Sab to Sd types with a drop from Sab to S0 types reaching about 27\%, consistent with that reported in \citet{Laurikainen2013, Ann2015, Buta2015, DiazGarcia2016}. The apparent increase to about 85\% at Sdm types should be taken with care due to small numbers in this morphological bin. The fraction of barred galaxies in terms of morphological types is not bi-modal, in contrast to \citet{DiazGarcia2016} that find a double-humped distribution in the Hubble sequence with a local minimum at Sbc types for the infrared S$^4$G sample \citep{Sheth2010}.

The fraction of barred galaxies as a function of \ms\ shows an incipient bi-modality with a minimum at $\log(\ms/\msun)\sim$9.9, contributed mainly by S0--Sb types on the high-mass region and by late Sbc--Sm types in the low-mass region. \citet{Nair2010b} in contrast, show a more clear bi-modality with the bar fraction falling steeply from the low mass region to the intermediate masses at about $\log(M_{\ast}/\msun)\sim$10.2, then rising slowly and having a plateau at higher masses. Other results \citep[e.g.][]{Masters2012} show also an incipient bi-modality. 

The fraction of barred galaxies in terms of $g-i$ colour shows no evidence of bi-modality in contrast to the results in \citep[e.g.][]{Nair2010b}. Instead, the bar fraction shows a decreasing trend with $g-i$ colour, though for redder colours the decrease tends to flatten. The number of galaxies with colours redder than 1.1 mag is very small, so that the reported bar fractions for these galaxies should be interpreted with caution. The contribution of S0--Sb morphological types is more important to red colours while late Sbc--Sm spirals are mainly contributing to the blue colours. The existence of bi-modalities in the stellar mass and colour distributions of barred galaxies suggest a link between the red and blue sequences and the origin and evolution of bars. However, our results do not show clear evidence of these.

In addition to the identification of bars, we also carried out a more refined classification of bars in terms of bar families following the Comprehensive de Vaucouleurs revised Hubble-Sandage system \citep[CVRHS;][]{Buta2015}. These results and a comparison with other morphological sources will be reported in a forthcoming paper of this series.

\subsection{Bright Tidal Features}

The identification of tidal features in a given sample depends on various factors, among them, variations in targeting strategy, detection methods, as well as differences in imaging depth (for example, tidal features may have a peak surface brightness at values $\sim$ 30 mag arcsec$^{-2}$; \citealt{Johnston2008}). We emphasize that the results presented here are based on a visual inspection of the more evident (bright) tidal features in the MaNGA DR15 sample, and that contrary to other authors, our visual inspection were based only upon the DESI $grz$ colour images without any further enhancing post-processing. 

Out of 4614 galaxies in the MaNGA sample, we find evidence of bright tidal features in 755 galaxies of all morphological types, corresponding to 16.4$^{+1.1}_{-1.0}$\% in the original sample and to 14\% in the volume-corrected sample.\footnote{Errors given are the range of possible percentages within 95\% confidence interval using the binomial exact confidence intervals method.} This volume-corrected fraction is considered as a lower limit on the true frequency of tidal features in the MaNGA DR15 sample. Although not directly comparable, our fraction is similar to that reported in \citet{Hood2018} (17 $\pm$ 2 \%) after a search for tidal features around galaxies by using $r-$band images from Data Release 3 of the DECam Legacy Survey 9 \citep[DECaLS;][]{Blum2016}.

\begin{figure}
\begin{center}
 \includegraphics[width=0.8\columnwidth]{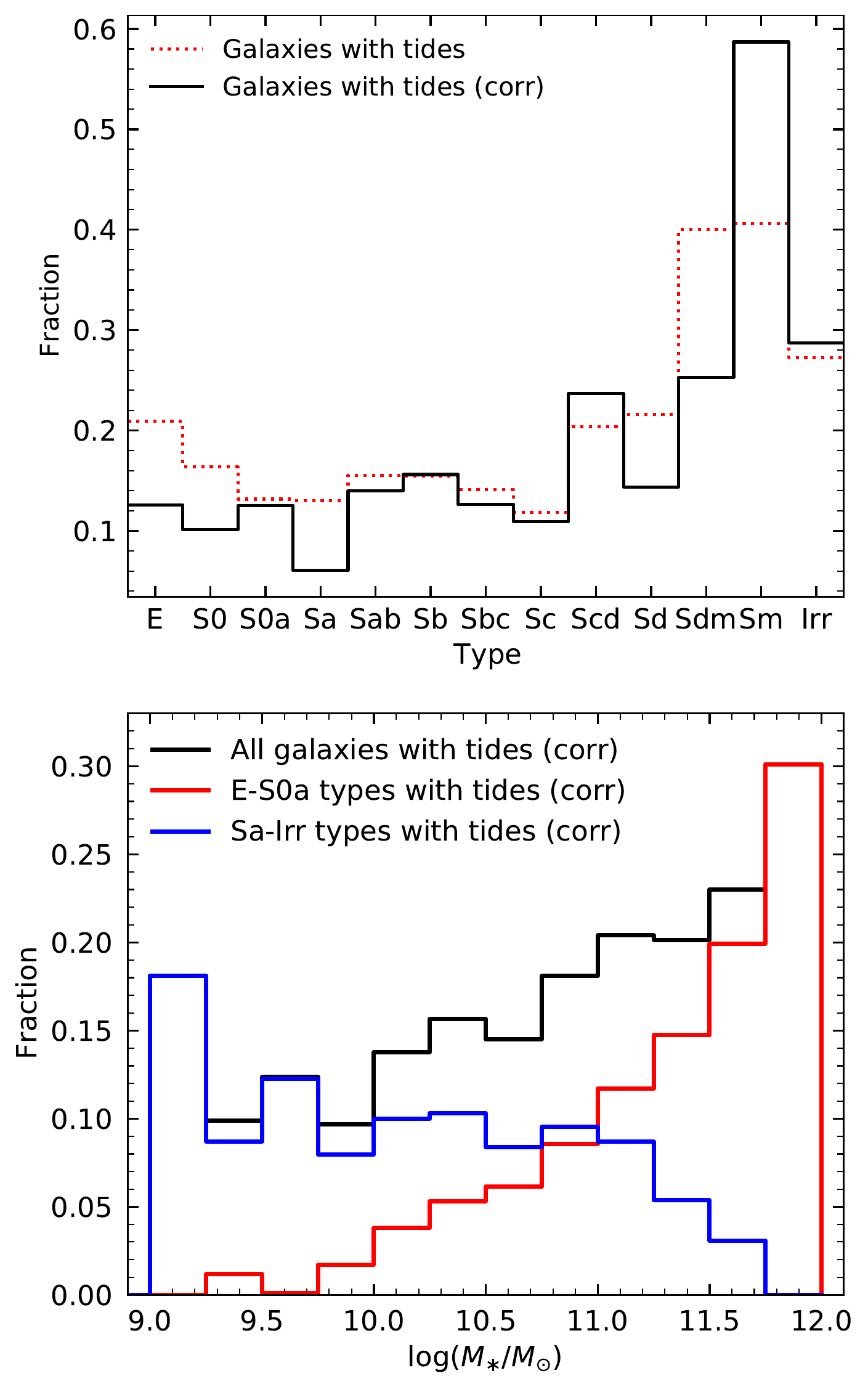}
 \caption{Fraction of galaxies with tidal debris as a function of morphology (upper panel) and stellar mass (lower panel), for the original (dotted-line) and volume-corrected (solid-black-line) samples. The sample of galaxies with tidal debris is split into E-S0a (red) and Sa-Irr (blue) types in the lower panel. 
 }
 \label{fig:tides_hist}
\end{center}
\end{figure}

Figure~\ref{fig:tides_hist} shows the volume-corrected fraction of tidal features in bins of morphological types (upper panel), and in bins of \ms\ (lower panel). The original fractions are presented as dotted lines for comparison. The upper panel shows that tidal features are present in galaxies of practically all the MaNGA morphological domain. Given the small numbers in the (Sd--Irr) bins, the reader should be cautious of not over-interpreting those apparent strongly increasing fractions.  

The fraction of tidal features in terms of \ms\ (lower panel) shows a distribution increasing monotonically from $\log(\ms/\msun)\gtrsim$ 9.4 to higher stellar mass bins, with an isolated maximum at the lower mass bin ($\log(\ms/\msun)\sim$ 9.1). When the distribution is split into two morphological groups (E--S0a in red) and (Sa--Irr in blue), (i) both groups contribute to the tidal fraction along the whole mass interval, (ii) an important contribution to the low mass end is coming, mainly from the Sa--Irr group, and (iii) the E--S0a group has an important contribution mainly to the high mass end. A fraction of the early type galaxies with tidal features are among the most massive galaxies.

\section{Comparison with other works}
\label{sec:compare}  

To explore the connection between a demographic view of the MaNGA survey and the inner physics of galaxies inferred from the IFU data analysis, requires of a detailed and uniform morphological classification of the MaNGA galaxies at its basis. We next carry out a comparison of our morphological results with other works considering subsets of galaxies in common and commenting on the implications, without considering any volume-correction.

\subsection{MaNGA DR15 Visual Morphology and Previous Classifications}

\citetalias{Nair2010} presented a detailed $g-$band visual classifications for 14,034 galaxies from SDSS DR4 in the redshift range 0.01 $< z <$ 0.1 down to an apparent extinction-corrected limit of $g<$ 16 mag. In addition to T-Types, they report the existence of various disk structures (bars, rings, arm multiplicity and other structures) and thus that work represents a valuable reference to compare our results. We find 1150 galaxies in common with the MaNGA DR15 sample and carry out a comparison under the following assumptions: (i) that any feature reported by \citetalias{Nair2010} has a visible counterpart in the DESI images, and that (ii) in case that \citetalias{Nair2010} do not report a feature, that does not imply that such feature may not be a visible feature in the DESI images.

\begin{figure*}
 \includegraphics[width=0.95\textwidth]{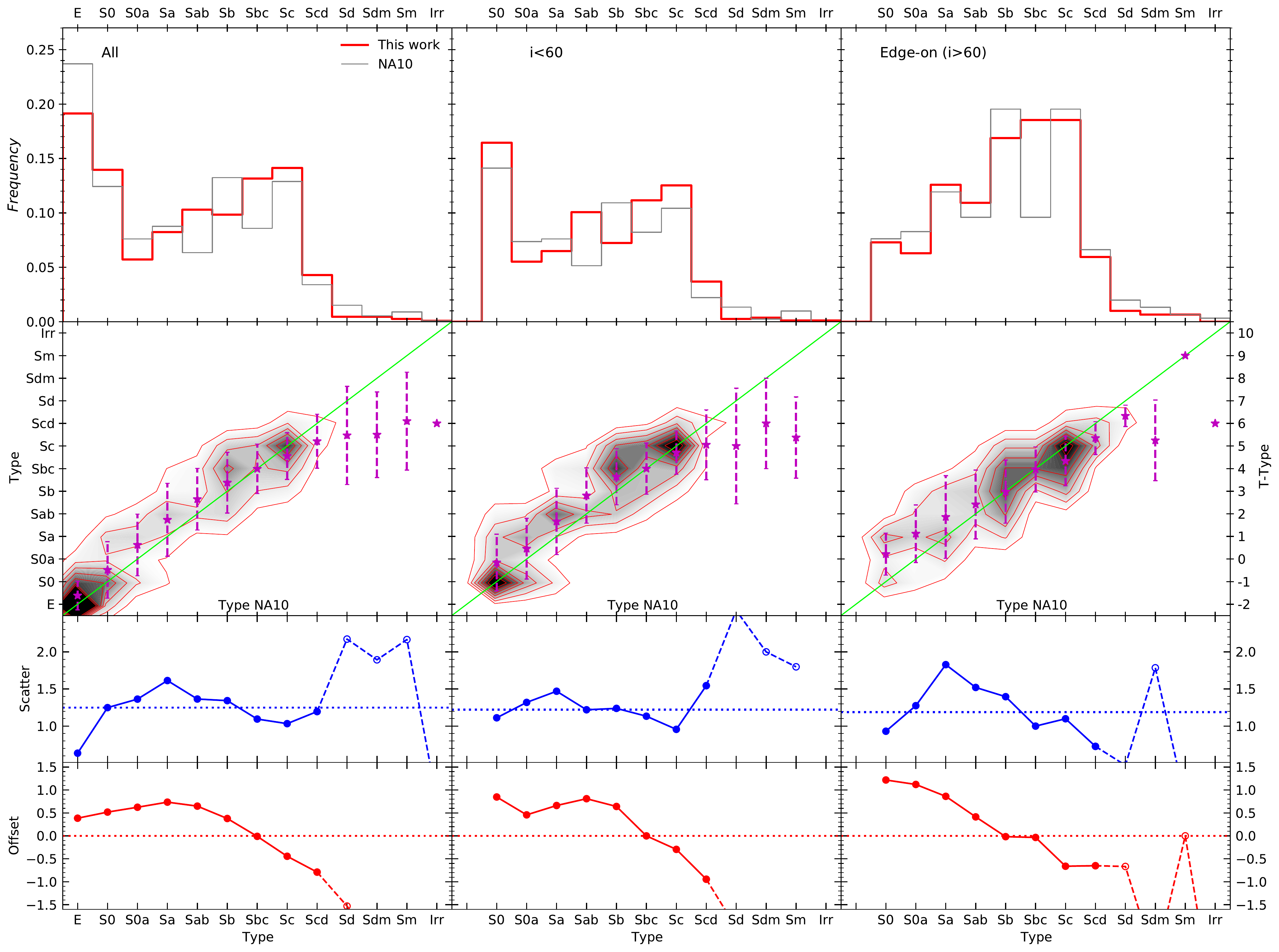}
 \caption{{\it Upper panels:} Frequency distributions of morphology from this work (red) and from \citetalias{Nair2010} (grey) classifications for the sample in common. This sample is split into galaxies with low ($i<$ 60; middle panels) and high inclination ($i>$ 60; right-hand panels). {\it Middle row panels:} \citetalias{Nair2010} classification vs. this work classification in 2D density histograms normalised to the total number of the sample. Isocontours are presented in lineal scale and the most external is formed with bins containing at least 0.9\% of the sample (10 galaxies). The one-to-one line is presented in green and the average with the standard deviation (magenta) for each morphology in \citetalias{Nair2010}. {\it Lower panels} show our relative offset of the average with respect to the one-to-one line (red), and scatter with respect to the average (blue) for a given morphological bin in \citetalias{Nair2010}. Dotted-lines (blue) are the corresponding median for all values of scatter for types $\leq$Scd, and the zero point reference for the offset (red). Later types (dashed and open circles) are omitted due to the low number of galaxies in these bins.}
 \label{fig:Histclass_compN10}
\end{figure*}

Figure~\ref{fig:Histclass_compN10} (upper panel) compares the relative frequencies of morphological types from our classification with that from \citetalias{Nair2010} (gray line). The upper panel is split into three sub-panels emphasizing possible inclination effects (middle panel; $i < 60^{\circ}$) and (edge-on galaxies; $i > 60^{\circ}$). 
The middle row presents a direct bin-to-bin morphological comparison of our assigned types (ordinate) with those of \citetalias{Nair2010} (abscissa); the green line shows the one-to-one case. The 2D density contours are normalised to the total number of galaxies and are presented in linear scale. Isocontours were estimated with bins of at least 0.9\% of the sample, meaning that at least 10 galaxies are taken into account from the total sample (1150 galaxies) in that bin. The lower panels show the scatter (the corresponding 1$\sigma$ from the mean in our classification given a T-type in \citetalias{Nair2010}) in blue and the relative offset (the difference between our numerical T-type mean and the corresponding value on the one-to one solid-green line) in red for a given reference morphological bin in \citetalias{Nair2010}. Notice that for types later than Sd, the number of galaxies is very small and were not considered in our comparison.

A global comparison show a median offset of b = 0.39 up to T-type $\leq$ Scd with a median scatter in our classification of 1.2 (dotted-blue line). When split by inclination, the low and high-inclination comparisons (middle and right panels) show a median offset of b = 0.64 with median scatter of 1.2, and median offset of b = 0.2 with a median scatter of 1.2, respectively if T-types up to Scd are considered. These results are considered in the range of those in reasonable agreement among classifiers \citep{Naim1995}.

The observed offset range, -0.9 $\lesssim$ Offset $\lesssim$ 0.7, has two behaviours. {\it (i)} A positive component in the E--Sb interval, where our morphological assignation offsets to later types for a reference type in this interval. These results suggest that in addition to the bulge component, we are identifying various disk structures in low and high inclination E--Sb galaxies in the DESI images as the presence of arms (and their winding and relative resolution into clumpy structures), compared to those identified in the $g$-band SDSS images. {\it (ii)}  A negative component for Sbc--Scd types, where our assignation offsets to earlier types for a reference type in this interval. The sense of this offset indicates that we are detecting a more prominent bulge contribution in our $r-$band DESI images, more sensitive to the light coming from an older stellar population, that makes our classification slightly earlier, compared to the reference classification in the SDSS $g$-band.

Our next comparison is with \citetalias{DominguezSanchez2021}, an updated version of the results reported by \citet{Fischer2019} for DR15 galaxies.
The \citet{Fischer2019} and \citetalias{DominguezSanchez2021} catalogues were generated based on a Convolutional Neural Network (CNN) algorithm presented by \citet{DominguezSanchez2018}, trained with two visually based morphological catalogues: Galaxy Zoo 2 and \citetalias{Nair2010}.  
\citetalias{DominguezSanchez2021} provide the most complete morphological classification of the MaNGA DR15 galaxies using Deep Learning models. Additionally, they complement the GZ2 classification with a T-type model and bar identification trained with the \citetalias{Nair2010} catalogue. Out of 4614 galaxies classified in the present work, 4609 galaxies are in common with \citetalias{DominguezSanchez2021}, making this a suitable comparison. Note that, although \citetalias{DominguezSanchez2021} propose different ways to select E and S0 galaxies, for this comparison we have considered only their T-type and S0 probabilities information, in order to include all galaxies. 

Some important aspects to highlight in this comparison are: (i) the high performance of their CNN model for intermediate T-types $<$ Scd and (ii) their E--S0 separation, providing a probability $P_{S0}$ of being S0 rather than E.
 
\begin{figure*}
 \includegraphics[width=0.95\textwidth]{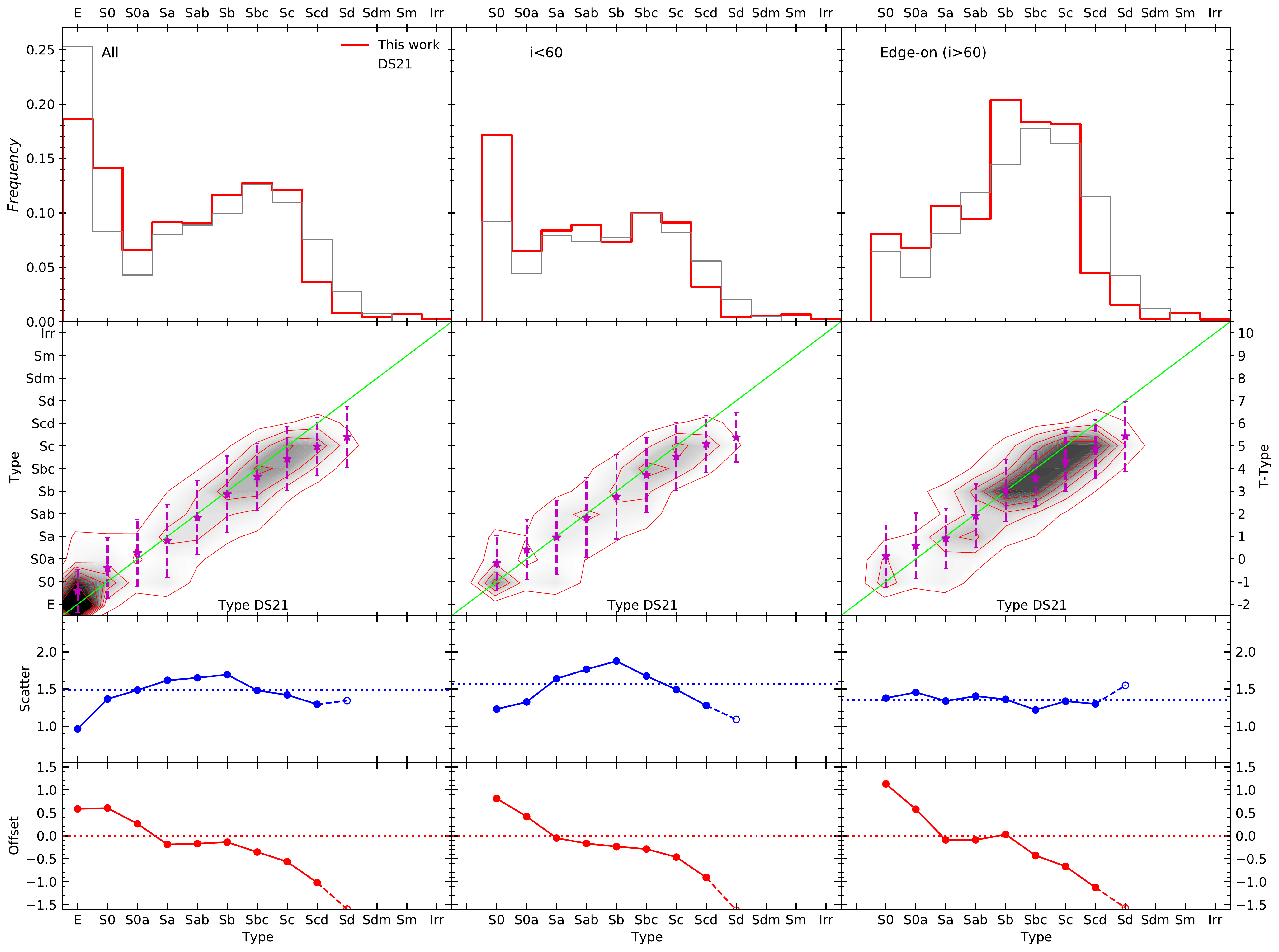}
 \caption{Same as in Fig. \ref{fig:Histclass_compN10} but for the \citetalias{DominguezSanchez2021} classification.
 }
 \label{fig:Histclass_compF19}
\end{figure*}

Similarly to Figure~\ref{fig:Histclass_compN10}, in Figure~\ref{fig:Histclass_compF19} we summarise the comparison of our results with those from \citetalias{DominguezSanchez2021}. Note that in the middle panels, the most external isocontour is formed this time by bins containing at least 41 galaxies. Notice also that no comparison is carried out for T-types later than Scd. 

A global comparison (left panel) show a median offset of b = -0.17 up to T-type $<$ Scd with a median scatter in our classification of 1.48. When split by inclination, the low-inclination panel shows a median offset of b = -0.2 up to T-type $\le$ Scd with a median scatter of 1.56 while the high inclination panel shows a median offset of b = -0.09 up to T-type $<$ Scd with a median scatter of 1.35. These results are also considered in reasonable agreement for independent classifiers \citep{Naim1995}.

The observed offset range, -1.0 $\lesssim$ Offset $\lesssim$ 0.7, shows very similar trend than in the previous comparison. This is somewhat expected since the images used to train the CNN algorithms were the colour composite $gri$ images from the SDSS database. This result emphasizes that any prediction from these algorithms depends directly on the quality of the training images and their association rules. Although in general, we find a similar trend to that found with Nair, here there are noticeable differences in our morphological assignation that are attributed to our combination of images and their post-processing, allowing us to identify in the case of S0 galaxies various inner and outer structures, and of outer structures and the bulge in the case of Scd and Sd edge-on galaxies.

\subsection{E and S0 separation}

\citet{Cheng2011} presented a light-based visual classification for early type galaxies that was further translated into a parameter space, bulge to total light ratio ($B/T$), SDSS axial ratio ($b/a$) and concentration index ($C$) with the goal to provide an automated classification scheme. We adopt that scheme to test our early-type classification also estimating both the completeness and purity achieved. To this purpose we adopt the $B/T$ values from \citet{Fischer2019}, while $b/a$ and $C$ values come from NSA \citep{Blanton2011}. Following \citet{Cheng2011} we recognize  objects in the region $B/T$ $>$ 0.5, $b/a$ $>$ 0.65 as candidates for the automated Elliptical sample.   
 
\begin{figure}
 \includegraphics[width=\columnwidth]{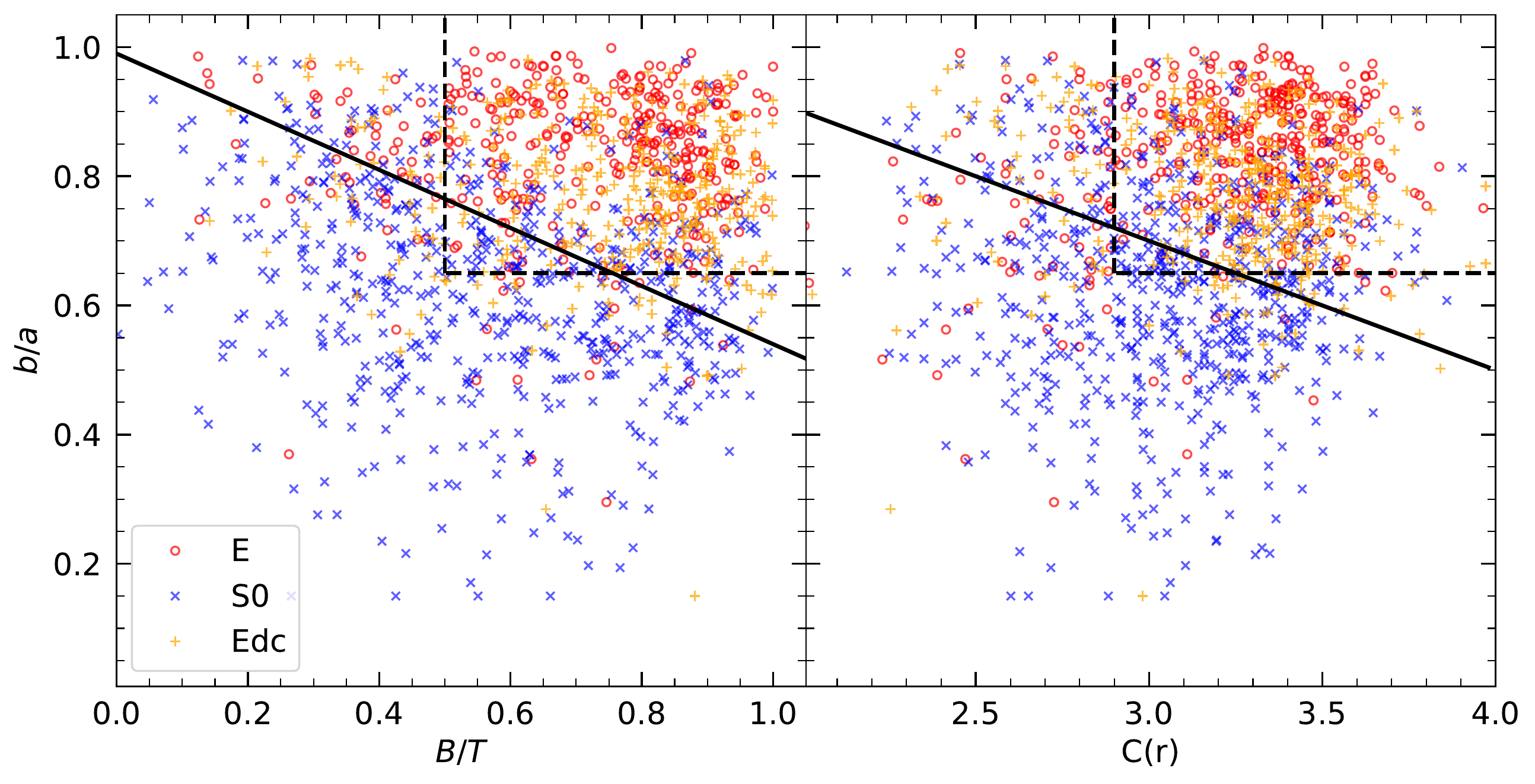}
 \caption{{\it Left panel:} the $b/a-B/T$ diagram for the early type galaxies E (red open dots), E with central discs (orange plus) and S0 (blue crosses) for our classification. {\it Right panel:} the same early type galaxies in the $b/a$ vs concentration ($C$) diagram. Dashed lines are the automatic separation proposed by \citet{Cheng2011} to separate bulge dominated from disk dominated galaxies. Solid-black lines show our proposed separation based on the Logistic Regression algorithm. Note that for this separation, we are considering Ellipticals with central disk as Ellipticals.}
 \label{fig:ES0_comp}
\end{figure}

Figure~\ref{fig:ES0_comp} shows the results of our classification for Ellipticals (red open dots), Ellipticals with inner disk-like features Edc (orange plus) and S0 (blue crosses) galaxies in the $B/T-b/a$ diagram. The plot also shows the boundary region (dashed-black lines) corresponding to the automated bulge classification. According to \citet{Cheng2011} the completeness is the percentage of all visual bulges that are recovered by the automated bulge criterion, while the purity is the percentage of all bulges that are truly visual bulges in the defined regions. We find a purity of 0.81 and completeness of 0.72, comparable to those reported by  \citet{Cheng2011} for bulge-dominated galaxies ({\it bulge class,} BC = 1; 0.73 in purity and 0.75 for completeness). Placing the results from \citetalias{DominguezSanchez2021} into this diagram, the corresponding results for purity and completeness are 0.89 and 0.60 respectively.

Figure~\ref{fig:ES0_comp} also shows an alternative separation region in the $B/T-b/a$ diagram for samples of galaxies containing only E and S0 types (i.e., not including Sa types). This is intended by a diagonal line (solid-black) devised after applying a Machine Learning algorithm of Logistic Regression, implemented in the Sklearn distribution \footnote{\url{https://scikit-learn.org}}, useful to maximising the number of E and S0 galaxies in each side of a predicted line. 

\begin{equation}
\label{eq:BT}
    b/a = -0.45 \times (B/T) + 0.99
\end{equation}

Following Eq.~\ref{eq:BT} we obtain a purity of 0.8 and completeness of 0.78.  The extension of the diagonal line to lower $B/T$ is justified from the results from detailed 2D decomposition image analysis (e.g. \citealt{Laurikainen2011} and \citealt{Graham2019}), showing that $B/T$ values in S0s extends to the region of the late-type spirals, covering the full $B/T$ range to near zero values. \citet{Laurikainen2011} report that S0 galaxies have a mean $B/T$ (in the $K-$band) ratio of 0.24 $\pm$ 0.11, which is significantly smaller than the average $B/T$ ($K-$band) = 0.6 reported in previous works. 
If we further place the \citetalias{DominguezSanchez2021} E and S0 results, by considering the same diagram alternative diagonal separation region (Eq.~\ref{eq:BT}), a purity of 0.88 and a completeness of 0.62 are estimated, where completeness is much lower compared to our results.

Since the automatic separation in \citet{Cheng2011} is based on structural parameters, we also propose to use the $C-b/a$ diagram including the concentration ($C$), another frequently estimated parameter. Figure~\ref{fig:ES0_comp} (right panels) shows our resulting E (red dots), Edc (orange dots) and S0 (blue dots) distributions in this diagram. We adopt $C$ $>$ 2.9, $b/a$ $>$ 0.65 for the automated bulge sample. Based on this diagram, we find a purity of 0.78 and completeness of 0.78, compared to 0.89 and 0.68, respectively for the \citetalias{DominguezSanchez2021} classification. Similarly to Eq.~\ref{eq:BT}, we propose a diagonal separation line based on Machine Learning,

\begin{equation}
\label{eq:C}
    b/a = -0.2 \times C + 1.3 .
\end{equation}

Following this line, our results are 0.76 in purity and 0.87 in completeness, compared to 0.86 and 0.73 respectively for the \citetalias{DominguezSanchez2021} classification. We conclude that these results are (i) statistically comparable to those obtained from the $B/T-b/a$ diagram and also (ii) comparable among other similar attempts.

\subsection{Bars}

We next carry out a comparison of our results on bars by considering two sources of morphological information; \citetalias{Nair2010} and the MaNGA-GalaxyZoo \citep{Willett2013} catalogues. For samples in common with both catalogues we compare the normalised distributions of morphological types and further discuss on the similarities and differences.  

\subsubsection{Comparison with Nair \& Abraham}
\label{sec:compBNair}

\begin{figure}
 \includegraphics[width=0.9\columnwidth]{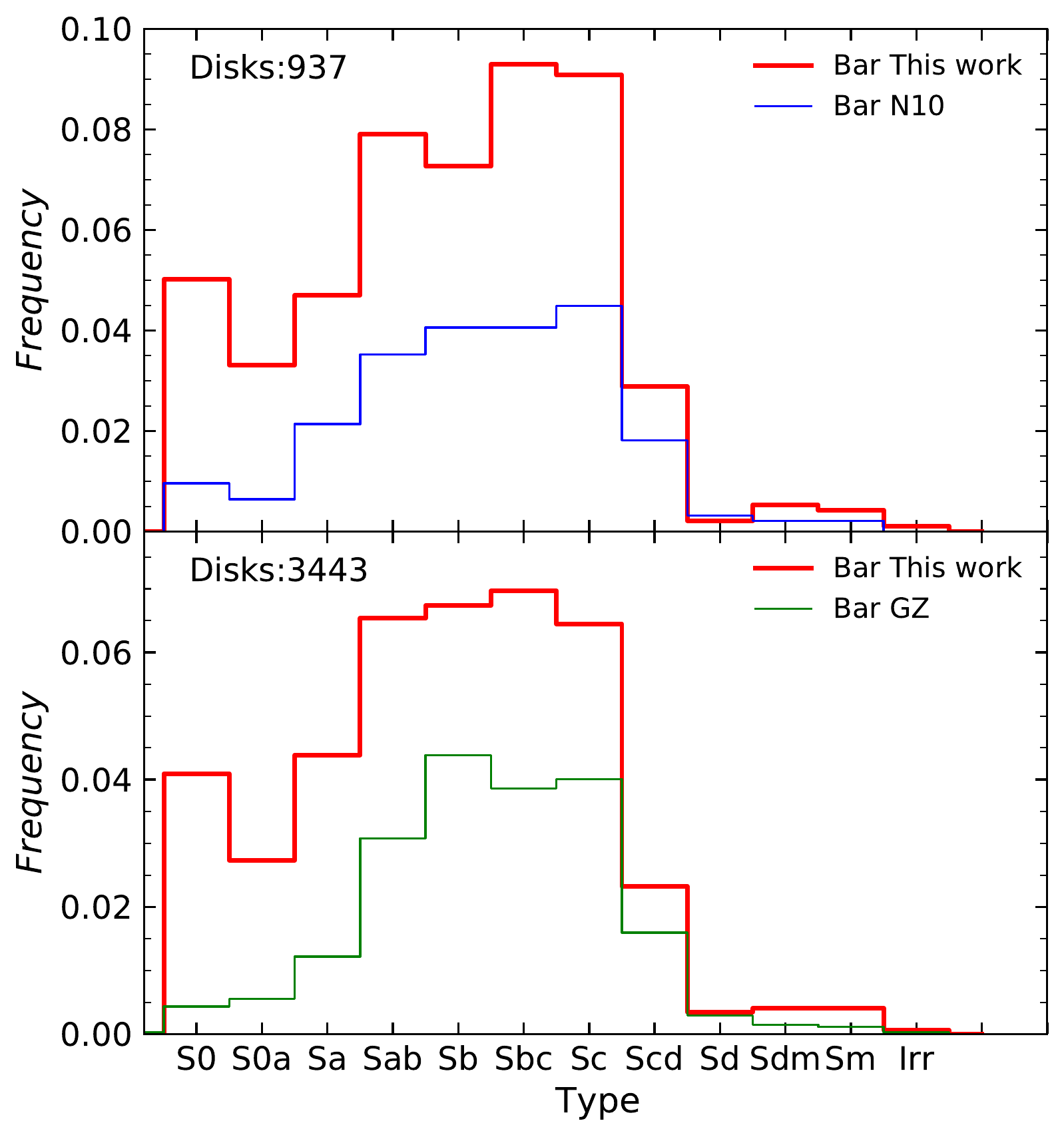}
 \caption{{\it Upper panel:} Comparison of our relative frequency distribution of barred galaxies as a function of morphology (solid-red line) with that of \citetalias{Nair2010} (solid-blue line) for the galaxy sample in common with those authors (galaxies with $i>70^\circ$ were excluded). The distributions are normalised to the 937 disks in common. {\it Lower panel:} Same as the upper panel but for the galaxies in common (3443) with the MaNGA-GalaxyZoo catalogue.}
 \label{fig:bar_comp}
\end{figure}

Figure~\ref{fig:bar_comp} (top panel) shows the relative frequency distribution of barred galaxies as a function of our morphological type, for a sample of 937 disk galaxies in common  with \citetalias{Nair2010}, normalised to this number.
Galaxies with $i>70^{\circ}$ were omitted to avoid possible bias for inclination effects. 

We detect bars in 513 (55\%) galaxies (475 with $i\le$ 70$^{\circ}$ and 38 with $i>$ 70$^{\circ}$). In contrast, \citetalias{Nair2010} report 200 (21\%) barred galaxies (194, with $i<$ 70$^{\circ}$ and 6 with $i>$ 70$^{\circ}$). Out of 200 galaxies classified as barred in \citetalias{Nair2010}, we have identified bars in 194 of them (97\%). However, for the sample in common, we further identify 286 barred galaxies not reported by \citetalias{Nair2010}, amounting to more than a factor of two bar identifications in our work. Notice that our distributions of barred galaxies as a function of \type\ is qualitatively similar to that of \citetalias{Nair2010}, except for S0 galaxies where our bar detection rate is much higher than in \citetalias{Nair2010}.

\subsubsection{Comparison with MaNGA-GalaxyZoo}

We carry out an alternative comparison of our results on bars with those reported in MaNGA-GalaxyZoo. According to the definitions and probability thresholds by \citet{Willett2013}, a bar can only be found in oblique disk galaxies, defined as $P_\emph{features/disk}$ > 0.430, that are not edge-on, $P_\emph{notedgeon}$ $\geq$ 0.715 and with a number of votes $N_\emph{notedgeon}$ $\geq$ 20. A probability to detect bars, $P_{Bar}$ was initially reported by \citet{Masters2011} to be $P_{Bar} >$ 0.5, mainly for strong bars. For comparison purposes we adopt the definitions in \citet{Hart2017} with thresholds of 0.2 $< P_{Bar} <$ 0.5 for weak bars (although with a probability of including other features than weak bars), and $P_{Bar} \geq$ 0.5 for strong bars.

Figure~\ref{fig:bar_comp} (lower panel) shows the relative frequency distribution of barred galaxies as a function of our morphological type, for a sample of 2735 disk galaxies with $i\le 70^\circ$ in common with MaNGA-GalaxyZoo.  Barred galaxies in this work (solid-red line), and MaNGA-GalaxyZoo bars (weak + strong; solid-green line) are normalised to the total number of disk galaxies in common. Similar to the comparison with \citetalias{Nair2010}, a trend of an increasing bar fraction from early types to a maximum near Sb-Sc types is appreciated, emphasizing also a fraction of bars in S0--Sa galaxies identified in our case but not in MaNGA-GalaxyZoo.

By considering $P_{Bar} \geq$ 0.2 in the sample in common, 680 (19.7\%) barred galaxies were found in MaNGA-GalaxyZoo ($i<70^\circ$). In contrast, we identify 1429 (41.5\%) barred galaxies for the sample in common. It is noticeable that aside of recuperating 631 out of the 680 galaxies classified as barred in MaNGA-GalaxyZoo, we have also identified a factor of $\sim 2$ more bars.

More recently \citet{Walmsley2021} have exploited the greater depth of DECaLS images, designing a new set of example images and more definite questions to improve sensitivity for the detection of bars and bar families in the GalaxyZoo platform, as \citet{Geron2021} already started to explore. In a forthcoming paper with the morphological results for the whole MaNGA DR17 sample at hand, a more detailed comparison of these results with our bar family identification will be presented.

\begin{figure}
 \includegraphics[width=\columnwidth]{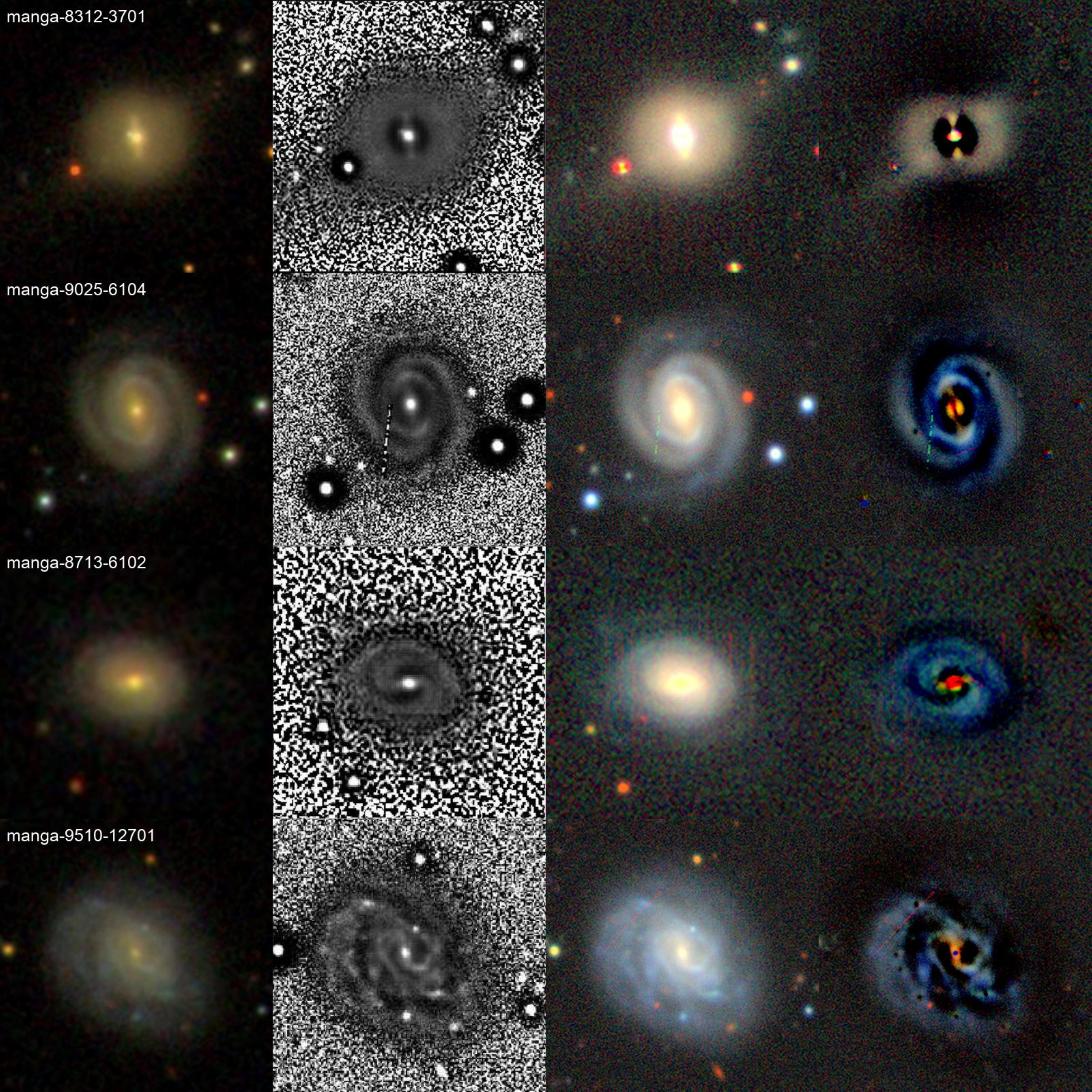}
 \caption{Examples of bar detection in galaxies in common with \citetalias{Nair2010} and the MaNGA-GalaxyZoo.
 From left to right, the \emph{gri} SDSS, the filtered-enhanced, the \emph{grz} DESI, and the residual DESI images as described in Sec~\ref{sec:steps}. First (manga-8312-3701) and row (manga-9025-6104) illustrates a strong/definite (B) bar for us, but not detected in \citetalias{Nair2010}. Third (manga-8713-6102) fourth row (manga-9510-12701) are example of a strong (B) bar in our case but not identified in MaNGA-GalaxyZoo.}
 \label{fig:bar_mos}
\end{figure}

Figure~\ref{fig:bar_mos} illustrates some examples of galaxies identified as barred according to our classification but that were not reported as barred in \citetalias{Nair2010} (first two rows) and MaNGA-GalaxyZoo (third and fourth bottom rows). The images are similar to those in Figure~\ref{fig:ES0S0a_mosaic2}. Notice how our filtered enhanced post-processing (middle-left column) in combination with the DESI residual image (right-most column) let us identify in each case a definite bar instead.

\subsection{Tides}

As part of their morphological evaluation on SDSS images, \citetalias{Nair2010} also identified interaction signatures, and classify them according to their appearance into disturbed, tidal tails, shells and bridges. The MaNGA-GalaxyZoo catalogue also reports merging galaxies, but without referencing to any specific tidal features. Since our identification does not make reference to any particular class of tidal features, but report instead only their presence or absence, a direct comparison with previous works is not possible. However, it is interesting to notice that out of 1150 galaxies in common with \citetalias{Nair2010} catalogue, we detect a total of 227 galaxies with bright tidal debris, while \citetalias{Nair2010} report only 78, 59 of them in common with us.

\begin{figure}
\centering
 \includegraphics[width=0.7\columnwidth]{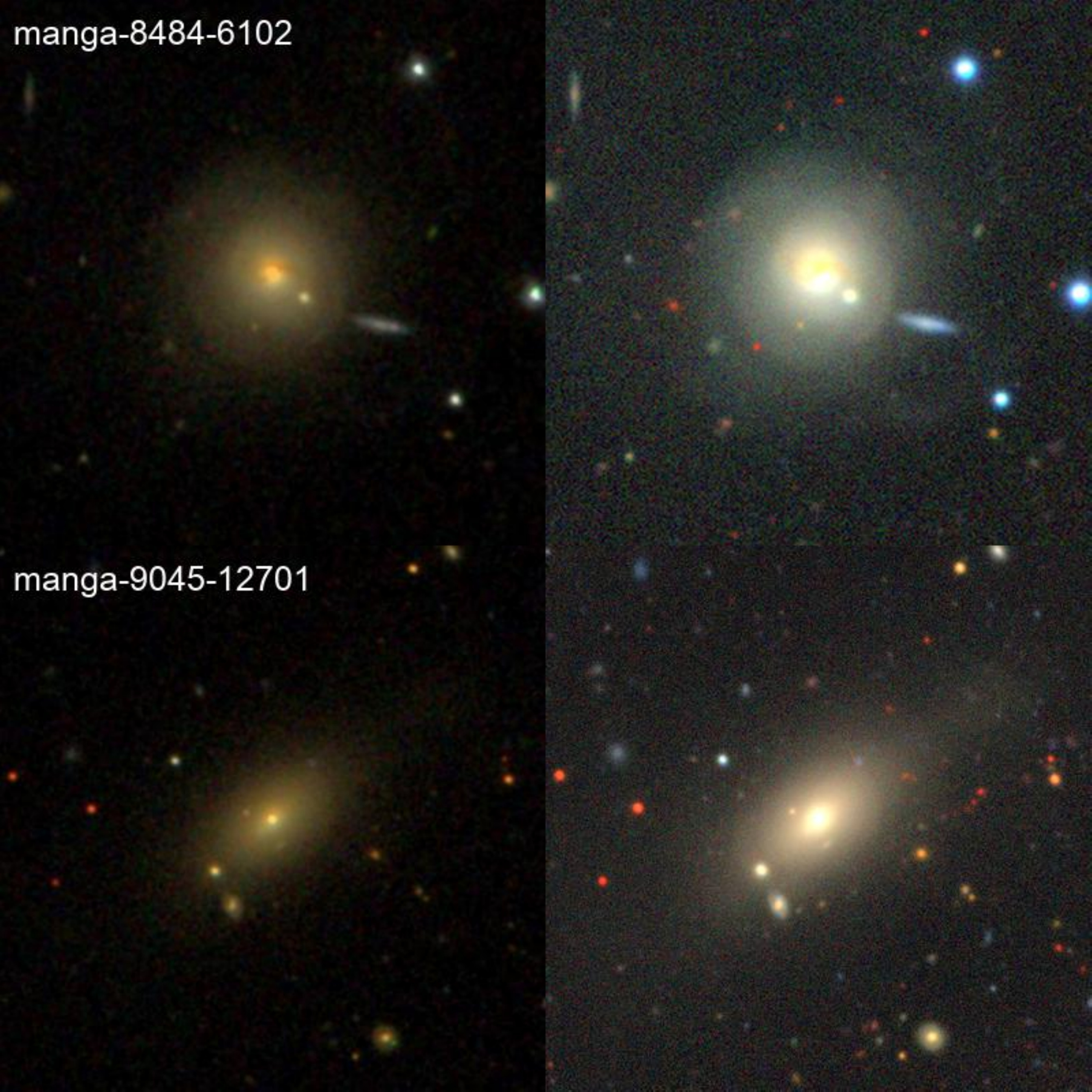}
 \caption{{\it Upper panel}: manga-8484-6102. {\it Lower panel}: manga-9045-12701. {\it Left}: $gri$ SDSS colour composite images. {\it Right}: $grz$ DESI colour composite images. Examples of galaxies with no tidal features as reported in \citetalias{Nair2010} but showing evidence of tidal features in the present work.}
 \label{fig:tides_mos}
\end{figure}

Figure~\ref{fig:tides_mos} illustrates some examples of galaxies in common with \citetalias{Nair2010} reported in that work as showing no evidence of tidal features. These same galaxies were identified, in contrast, as showing clear evidence of tidal features in the present work, emphasizing the advantages of the deeper images of the DESI survey. 

\begin{figure*} 
 \includegraphics[width=0.8\textwidth]{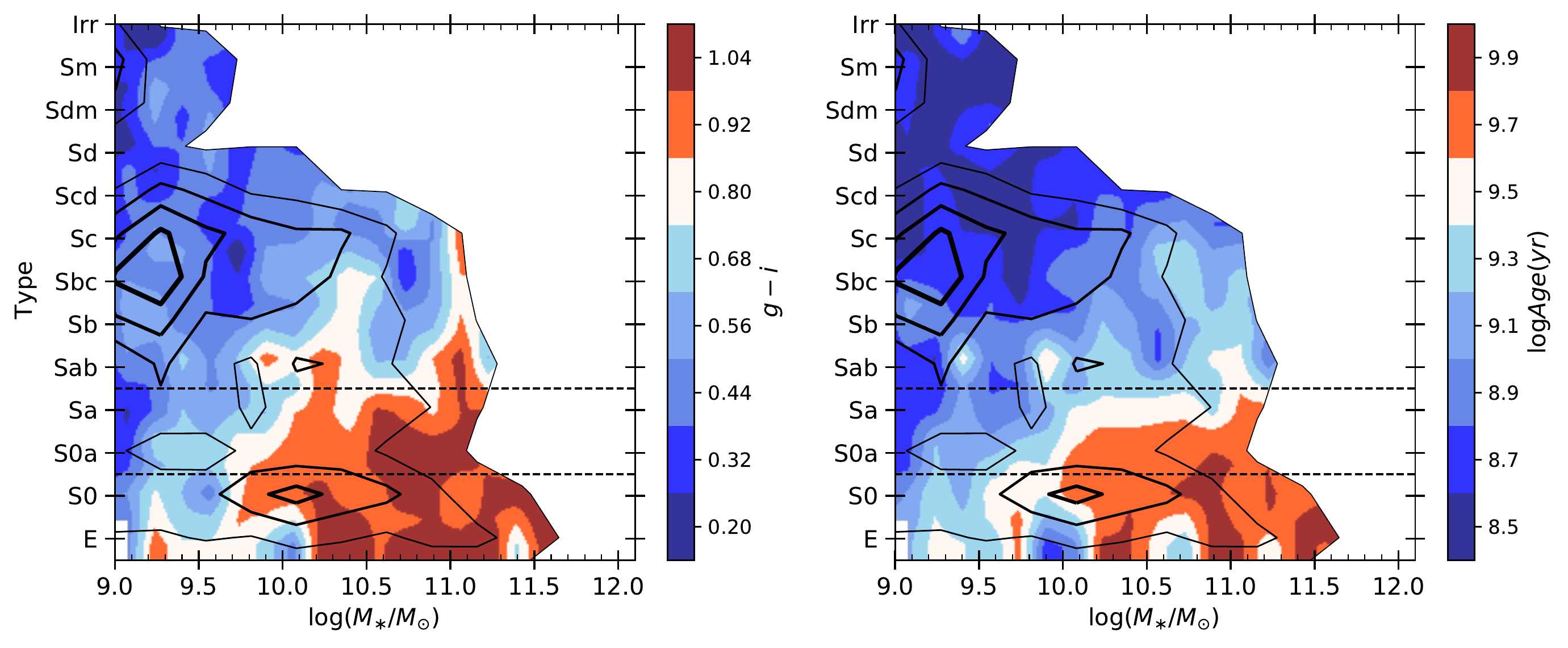}
 \caption{Bi-variate distributions in the \type--\ms\ diagram (isodensity contours and dashed lines are the same as right-hand panel in Figure \ref{fig:mass_morpho}) but coloured by the dust-corrected $g-i$ colour (\textit{left panel}), and by the luminosity-weighted age (\textit{right panel}) as indicated in the colorbars, respectively.  Colours within roughly $g-i=0.8\pm 0.05$ mag and ages within roughly 2.5--4.0 Gyr (reddish white colour in both panels) can be considered as transitioning between red and blue galaxies, and between old and young galaxies, respectively. Galaxies more inclined than $70^\circ$ were excluded. 
 }
 \label{fig:AGE}
\end{figure*}

\subsection{Final remarks}

We have compared the results of our visual morphological classification with various sources of morphological information based on other independent set of images. The classifications of \citetalias{Nair2010}, GZ and \citetalias{DominguezSanchez2021} rely on the use of the original SDSS images ($g-$band) and their corresponding colour composite ($gri$)  images. The SDSS and DESI data sets have different characteristics in terms of depth, PSF, and instrumental effects. Thus it is expected that part of the differences found in the classification  may be related to the different properties of the images. On one side, the deeper images from the DESI survey were able to confirm  the already definite and brighter features identified by the SDSS image survey, to mention, the presence of prominent bulges, bright spiral arms and their flocculency, bright bars and bright tidal features as relics of recent minor and major mergers. On the other side, we gained additional morphological information useful for the identification of intrinsically fainter features like faint bulges, weaker spiral arms in face-on and edge-on galaxies, weak bars and their surrounding structures, and faint tidal features not visible in the SDSS imaging. In this sense, the DESI images represents a reliable set of images for a detailed morphological analysis. 

Another component behind the differences found in our morphological evaluation is related to our post-processing of both the SDSS and DESI r-band images and the addition of the DESI residual images allowing us for a more thorough visual inspection of each MaNGA galaxy in the DR15 sample. We were able to identify various disk structures difficult to recognize in RGB and non-processed SDSS images, like inner structures against a dusty and clumpy background disk, faint and smaller bars as well as faint structures in the outskirts of galaxies at various inclination angles.

\section{Discussion}
\label{sec:Discussion}

\subsection{Colour and age distributions as a function of morphology and mass}
\label{sec:color-age}

The bi-modality in the \type--\ms\ diagram shown in the right panel of Figure \ref{fig:mass_morpho} resembles roughly the well known bi-modality in the colour--\ms\ diagram. Therefore, if we separate galaxies into LTGs and ETGs, the bi-modality in the respective colour--\ms\ diagram would tend to disappear because LTGs are mostly blue and ETGs are mostly red. This suggests that the colour--\ms\ bimodal distribution is mainly consequence of the \type--\ms\ distribution.
The above has been shown for a large sample of SDSS galaxies in \citet[][]{Schawinski+2014}.

In the left panel of Figure \ref{fig:AGE} we reproduce the isodensity contours of Figure \ref{fig:mass_morpho} corresponding to the bi-variate distribution of the MaNGA galaxies in the \type--\ms\ diagram ($i<70^\circ$) but the colours now indicate the corresponding photometric $g-i$ colour of galaxies given their morphological type and \ms.\footnote{The K-corrected to $z=0$ $g-i$ colours were taken from the NSA catalogue. We corrected these colours by the Milky Way extinction (taken also from the NSA catalogue), the internal dust attenuation using the $A_V$ coefficients adquired from the GALEX-SDSS-WISE Legacy Catalogue \citep{Salim2018} and the \citet[][]{Cardelli+1989} extinction law, and an evolution factor according to \citet{Dragomir2018}.} 
The segregation is clear. The LTGs (Sab--Irr) are mostly blue, $g-i\lesssim 0.75$ mag, with a small fraction of Sab--Sc galaxies more massive than $\sim 10^{10}$ \msun\ transiting to the red sequence. The ETGs (E--S0) are mainly red, $g-i\gtrsim 0.85$ mag, but with a transition to blue galaxies at the lowest masses, $\ms<5\times 10^{9}$ \msun, where their abundance is very small. 
Unlike LTGs and ETGs, notice that galaxies with intermediate morphological types, the ITGs (S0a--Sab) as established in \S\S \ref{sec:mass_Hubble}, present a bi-modal colour distribution consisting of red galaxies at high masses and blue galaxies at low masses, with a transitioning region across the  $9.7\lesssim\log(\ms/\msun)\lesssim 10.5$ mass range, the ``green valley'' (roughly the white colour in Fig. \ref{fig:AGE}). 
This green valley zone, an approximately diagonal region in the \type--\ms\ diagram, extends beyond the ITG region, towards later types (Sab--Sb) on the massive side, and towards earlier types (S0--E) on the low mass side. This zone, in general, is poorly populated and it separates the two densest regions in the \type--\ms\ diagram: the cloud of low and medium mass late-type objects, which are the bluest ones ($g-i\lesssim 0.6$ mag), and the sequence of medium and high mass early-type objects, which are the reddest ones ($g-i\gtrsim 0.9$ mag).

Our results confirm those of \citet[][]{Schawinski+2014} but extend them showing a colour bimodality for galaxies of intermediate morphologies, those of types S0a--Sa. It is for these galaxies that the bi-modal distribution in the colour--\ms\ diagram is evident, including the corresponding mass-dependent transition region called the green valley. 

The global $g-i$ colour is a proxy of only the recent ($\lesssim 1-1.5$ Gyr) mean SF history (SFH) of galaxies, while morphology is more related to the dynamical history of galaxies. From the spectral information provided by MaNGA IFS observations, we can calculate quantities related directly to the complete SF history of the galaxy. 
Here, we calculate the integrated (within the effective radius) \textit{logarithmic} mean luminosity-weighted age, $\age=\langle\log t\rangle_{\rm lw}$, of each MaNGA galaxy using Pipe3D \citep[][]{Sanchez2018}. 
\age\ is sensitive to relatively late SF activity and very recent SF episodes, so that we prefer to use it instead of mass-weighted ages.
In the right panel of Figure \ref{fig:AGE} we show the same as in the left panel but for $\age$ instead of $g-i$ colour. 
We define galaxies dominated by old stellar populations as those with \age$\ge 4$ Gyr and galaxies dominated by young populations, or with non-dominant but very recent events of active SF, as those with 
\age$< 2$ Gyr.\footnote{ Very young stellar populations contribute more to the mean age when the averaging is logarithmic than when it is arithmetic. Therefore, in galaxies with very young stellar populations, the \age{} computed logarithmically can be much lower than when computed arithmetically. }
Galaxies with ages between these values (white region
and part of the cyan region in the right panel) can be considered as intermediate-age galaxies, the ``green valley'' in age.

\begin{table}
  \begin{center}
    \caption{Fractions of galaxies in each morphological group according to their luminosity-weighted ages.}
    \label{table:age-fractions}
    \begin{tabular}{c|c|c|c|c|c|c}
      \hline
      \textbf{Morph.} & \textbf{Group}  & \multicolumn{3}{c|}{\textbf{Fraction within}} & \textbf{mean} & \textbf{median} \\
      \textbf{Group} & \textbf{\%} & \multicolumn{3}{c|}{\textbf{the Group}}  & \textbf{\ms$^a$} & \textbf{\ms$^a$} \\
      & & {Young} & {Interm.} & {Old}& &  \\
      \hline
ETG & 18.9 & 0.237 & 0.179 & 0.584 & 3.9 & 1.3 \\
ITG & 14.1 & 0.420 & 0.257 & 0.323 & 2.2 & 0.4 \\
LTG & 67.0 & 0.846 & 0.114 & 0.040 & 1.1 & 0.3\\
      \hline
      \multicolumn{3}{l|}{$^a$\ms\ is given in $10^{10}$\msun\ units}
    \end{tabular}
  \end{center}
\end{table}

The segregation of galaxies in late-type/young and early-type/old galaxies seen in the right panel of Figure \ref{fig:AGE} is strong. 
Table \ref{table:age-fractions} shows the fractions within each galaxy group (ETG, ITG, and LTG) that have young, intermediate and old luminosity-weighted ages. Within the LTG group (Sab--Irr), the age as a function of \ms\ follows roughly a gradual distribution, with the less massive being very young and the most massive having intermediate ages on average. {Also, as \type\ is earlier, \age\ is larger, though rarely larger than 4 Gyr.}
For the ETG group (E--S0), the \age\ distribution is also roughly gradual throughout the mass, with the massive being the oldest galaxies and the less massive having intermediate luminosity-weighted ages, or even young. For the ITG group (S0a--Sa), their distribution and range of ages is broad, with the most massive being old, 
the less massive being young, 
and a clear mass-dependent transitioning zone in the  $9.7\lesssim\log(\ms/\msun)\lesssim 10.7$ mass range. Therefore, the ITGs present a roughly bi-modal mass-dependent distribution in the $\age$--\ms{} plane, albeit with a notable fraction of galaxies in the intermediate age region.

The distribution of \age\ in the \type--\ms\ diagram resembles the distribution of attenuation-corrected $g-i$ colour in the same diagram. However, the distributions are smoother for \age\ than for the $g-i$ colour. As mentioned above, \age\ is expected to provide a more robust characterization of the SF history of galaxies than colour and it is less resilient to dust. 
Thus, since \age\ is a better proxy of the SF history of galaxies than colour, demographic studies in terms of \age\ can provide better clues about the SF ageing and quenching processes, and its connection with the morphological transformation of galaxies, in the spirit of those who use colours and SFRs \citep[e.g.,][]{Schawinski+2014,Tacchella+2019}. 
In \S\S\ \ref{sec:implications}, we discuss some preliminary interpretations of our results.

\subsection{Fractions of galaxies with bars and tides as a function of morphology and mass}

\begin{figure}
 \includegraphics[width=\columnwidth]{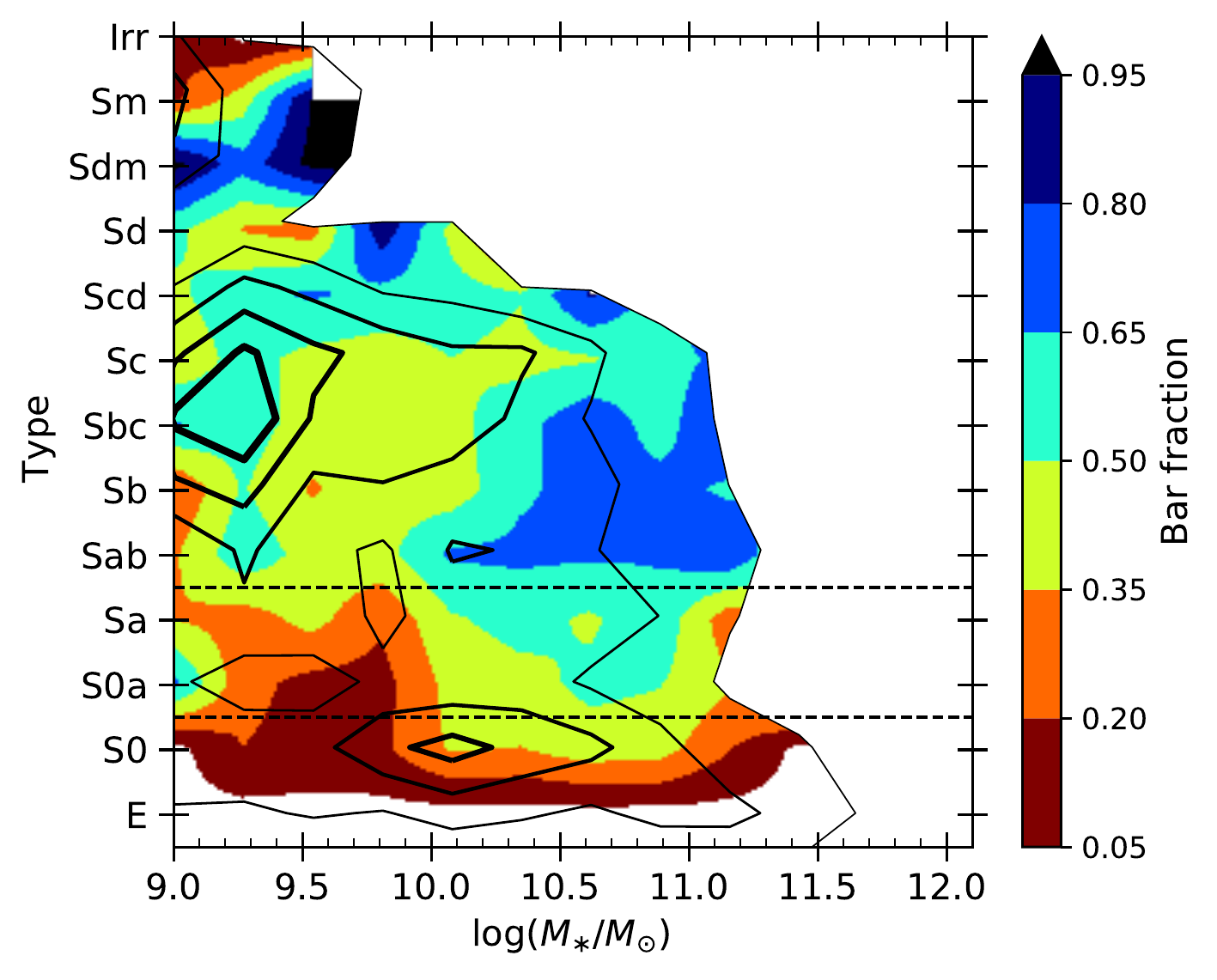}
 \caption{Bi-variate distribution in the \type--\ms\ diagram, similar to Figure \ref{fig:mass_morpho} but coloured by the bar fraction distributions (volume-corrected) as indicated in the colorbar.
 }
 \label{fig:Bar_massT}
\end{figure}

In \S\S \ref{sec:bars} we have presented the distributions of the fraction of MaNGA galaxies with bars as a function of \type, \ms, and dust attenuation-corrected $g-i$ colour. Here, we discuss on the trends of the bar fraction along the
joint \type\ and \ms\ distribution, similar as in the previous subsection for colour and age.  
Figure~\ref{fig:Bar_massT} shows the bi-variate distribution in the \type--\ms\ diagram (the isodensity contours are as in Figure~\ref{fig:mass_morpho}), but the colours indicate now the bar fraction, following the colorbar in the right. This figure complements the information on bar fractions provided in Figure~\ref{fig:Bar_fam} and Table~\ref{tab:mass}.  The bar fractions in one of the two densest regions in this plot, the cloud of low and medium mass late-type galaxies, are around $0.5\pm 0.15$, while the bar fractions in the other densest region, the sequence of medium and high mass early-type galaxies, are mostly lower than $\sim 0.35$.  The galaxies with the highest fractions of bars, $\gtrsim 0.7$, are the relatively massive Sab--Sbc spirals  and the low-mass Sdm--Sm spirals, but in both cases the frequency of these galaxies are low.

Interestingly, the largest variation in the bar fraction  in Figure \ref{fig:Bar_massT} is along the diagonal region that goes from low-mass S0s (fractions of $\sim 0.1$) to high-mass Sab--Sb spirals (fractions up to $\sim 0.8$). This region roughly corresponds to the transition zones in the $g-i$ colour--\ms\ or \age--\ms\ diagrams (see previous subsection), and are the least populated  regions in these diagrams as well in the \type--\ms\ diagram (see the isodensity contours in Fig.~\ref{fig:Bar_massT}). 
In general, the fraction of barred galaxies increases at all masses from S0 to Sab types (see also Fig. \ref{fig:Bar_massT}), but for later types, the fraction weakly depends on the morphological type.

\begin{figure}
 \includegraphics[width=\columnwidth]{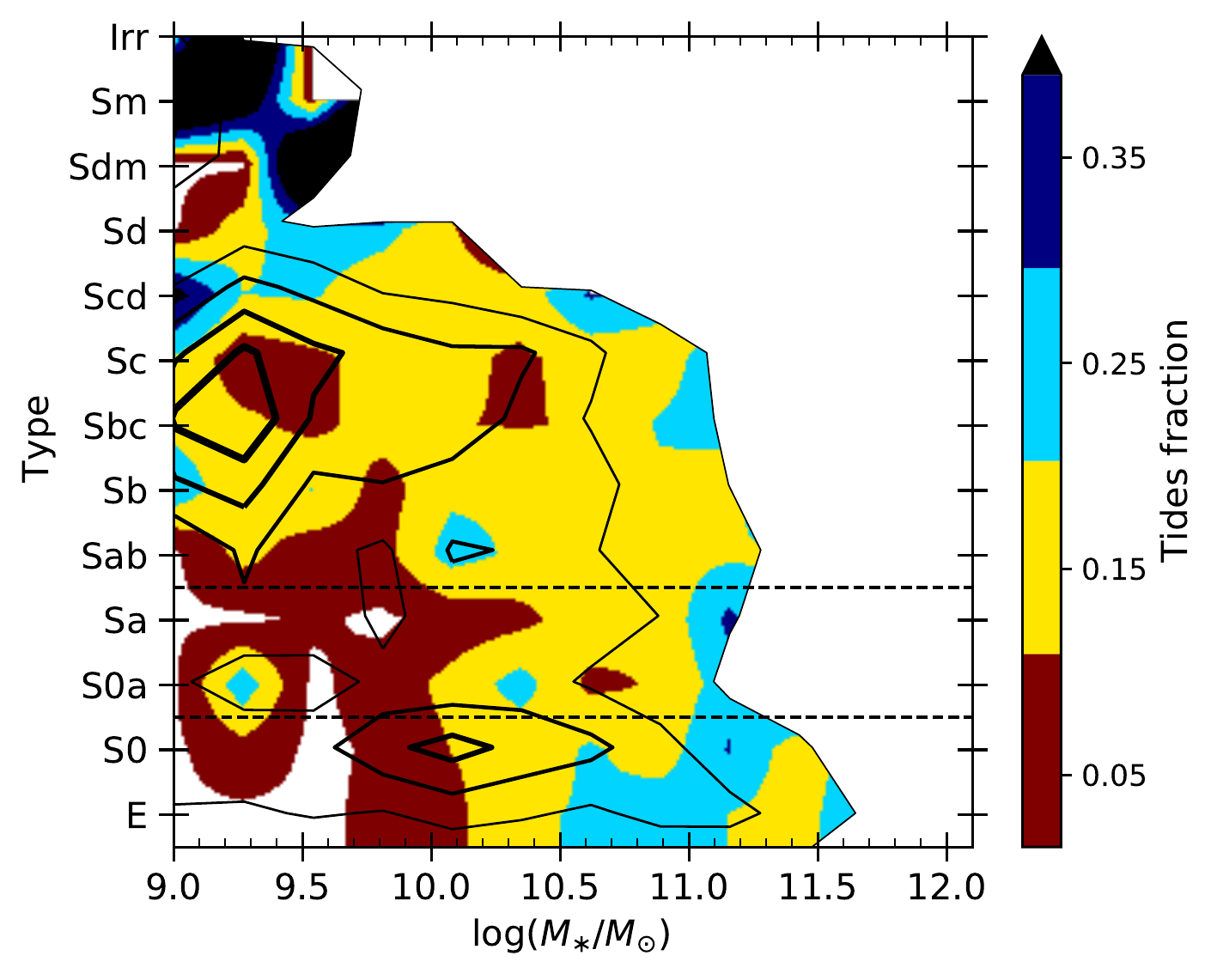}
 \caption{Bi-variate distribution in the \type--\ms\ diagram (similar to Figure \ref{fig:mass_morpho}), but coloured by the tides fractions (Volume-corrected) as indicated in the colorbar.
 }
 \label{fig:Tides_massT}
\end{figure}

Figure~\ref{fig:Tides_massT} shows the bi-variate distribution in the \type--\ms\ diagram (the isodensity contours are as in Figure~\ref{fig:mass_morpho}), but the colours indicate now the tidal features fraction, according to the colorbar on the right. This figure complements the information on tidal features provided in Figure~\ref{fig:tides_hist}. Overall, as already discussed, the fraction of galaxies with bright tidal features is low but there is a trend of increasing this fraction as the galaxies are of later types and more massive. For galaxies in the dense populated cloud of Sb--Scd spirals less massive than $\sim 3\times 10^{10}$ \msun\ in the \type--\ms\ diagram (corresponding also to the blue/young cloud), the tides fraction is small. For galaxies in the other populated region, that of E--Sa in the $\sim 0.5-10.0\times 10^{10}$ \msun\ (corresponding also to the red/old sequence), the tides fractions increases with \ms, but even for the most massive E--Sa, this fraction does not overcome $\approx 0.3$. In general, for each morphological type, the higher tides fractions are seen for the most massive galaxies. 

If the absence of tidal features is due the isolation of the galaxy, then the fact that the tides fraction in low-mass ETGs is negligible, as seen in Figure~\ref{fig:Tides_massT}, it would imply that these galaxies are mostly very isolated. Indeed this seems to be the case, see e.g., \citet[][and more references therein]{Lacerna+2016}. In the other extreme, the galaxies with the highest tides fractions are the low-massive Sdm--Irr ones. These are blue and young galaxies (see Fig. \ref{fig:AGE}), which probably are in their early phases of formation by gas and (some) flybys.

\subsection{Preliminary interpretation of our results}
\label{sec:implications}

According to our results, the vast majority of LTGs have low values of \age{ } (Table \ref{table:age-fractions}), that is, they contain an important fraction of young stellar populations, which implies that these disc galaxies  have been star-forming galaxies, and thus their bluer colours; the lower the mass the younger the galaxy, on average. However, \age\ gradually increases the larger \ms\ and  for earlier morphological types; for the less abundant LTGs with $\ms\gtrsim 5\times 10^{10}$ \msun{} and \type\ earlier than Sc, their \age\ values may exceed $\sim 2$ Gyr (Fig. \ref{fig:AGE}) 
These LTGs can be interpreted as in the process of quenching,
but the quenching timescales for them appear to be long, since most of them are not older than \age$=4$ Gyr, when ages $>4$ Gyr are typical of massive earlier-type galaxies (see Fig. \ref{fig:AGE}), which are quiescent or passive (quenched). 
The above is consistent with slow quenching mechanisms of SF for the massive LTGs, such as: (a) halo mass quenching (e.g., \citealp{DekelBirnboim2006,GaborDave2015}; recall that galaxies more massive than $5\times 10^{10}$ \msun\  reside in haloes more massive than a few $10^{12}$ \msun, where this mechanism is already effective).
(b) AGN feedback from an accretion disk that it is heating and/or ejecting the gas from the galaxy.   
It has been suggested in the literature that bars are a way to activate and/or fed AGNs in addition of major mergers and/or other disk instabilities; interestingly enough, our results show that the bar frequency in massive LTGs is high, see Fig. \ref{fig:Bar_massT}.
(c) Morphology quenching (e.g., \citealp{Martig+2009,Fang+2013}; the bulge-to-total ratios of most Sbc--Sab galaxies are higher than 0.25-0.30 and so, by this mechanism, such bulges can produce some disc stabilization that inhibits active SF).

As for the ETGs, which are spheroid-dominated, most of these that are more massive than $\sim 2\times 10^{10}$ \msun\ are old, implying that the SF has strongly declined long time ago and no recent bursts of SF have occurred. Early formation and rapid quenching mechanisms are expected for those galaxies, where the quenching mechanisms are likely to be associated to their morphological transformation through wet major mergers and disk instabilities. 
These processes lead to compaction and strong bursts of SF that consume the gas \citep[e.g.,][]{Hopkins+2008,Barro+2013,DekelBurkert2014}, as well as strong AGN/QSO and Supernova feedback that  will heat and expel the gas \citep[e.g.,][]{Granato+2004,Sijacki+2007,Somerville+2008,Vogelsberger2014}. At intermediate and low masses, the environment-driven quenching mechanisms (e.g., ram-pressure and tidal striping, strangulation, etc.) can also be relevant \citep[e.g.,][]{Kauffmann+2004,Peng+2010,Schawinski+2014}. Note that these mechanisms are expected to lead also to morphological transformation \citep[e.g.,][]{GunnGott1972,Moore+1996,Abadi+1999,Bekki+2002,Aragon-Salamanca+2006}. However, as \ms\ is lower there is an increasing fraction of ETGs with intermediate values of \age, or even low values (Table \ref{table:age-fractions}), though not as low as in the case of low-mass LTGs (see Fig. \ref{fig:AGE}). This population of ETGs is partially associated to the Blue Star-forming and Recently Quenched Early-type galaxies, for which rejuvenation processes \citep[][]{Thomas+2010} by gas infall or late gas-rich mergers were proposed in \citet[][]{Lacerna+2016,Lacerna+2020}. In fact, the $Age_{\rm mw}/\age$ ratio of these galaxies is large, which implies that they did not form so late but they contain some (small) fractions of very young stellar populations.  
Finally, for the ITGs, the \age\ distribution in the $\sim 0.5-5\times 10^{10}$ \msun\ mass range is roughly bimodal (Fig. \ref{fig:AGE}), with a large fraction of them in the intermediate age region (or green valley in the color--\ms\ diagram), showing that the quenching time-scales for these galaxies with a significant bulge are slow (quenching mechanisms like those driven by morphology, halo mass, and environment; see above). For the most massive S0--Sa galaxies, $\age>4$ Gyr, that is, they are mostly quenched, while for the lest massive, $\age\lesssim 2$ Gyr, which imply some rejuvenation processes.

We conclude that the SFH of galaxies and their quenching processes are closely related to the morphology and morphological transformation of galaxies, respectively. However, our results show that these cannot be the only factors. Other galaxy properties and evolutionary processes also play a role, such as the halo mass, the AGN feedback strength and its relation to the bars, and the local environment. 

In a forthcoming paper we will study in more detail these questions, taking into account several galaxy global properties (\ms, \type, colour, SFR, ages, metallicities, etc.) as well as radial gradients of some of these properties and the environment.

\subsection{The MaNGA galaxies that are difficult to classify}
\label{Sec:diffuse}

In Section \ref{sec:morpho-classification} we mentioned that $\approx 8\%$ of the galaxies in the MaNGA DR15 sample were difficult to classify. It is important to understand the nature of these galaxies that were flagged as ``unsure''. These cases refer to relatively isolated and intrinsically faint or diffuse objects, some of them barely showing signs of an apparent disk (S-like) at different inclinations, and a few of concentrated appearance with very few resolution elements. Notice that some faint dwarf/irregular galaxies may have entered in this category. However, cases like pre-mergers (binary almost fusing nuclei), mergers (two or more galaxies already collided plus tidal signatures) and bright dwarf galaxies (when recognised) were omitted from this category as much as possible. Our results indicate that 355 MaNGA galaxies in the DR15 sample were identified as ``unsure'' objects.
\footnote{Notice that with the help of our image post-processing, a very crude classification was in any case attempted for these objects.}

\begin{figure}
 \includegraphics[width=\columnwidth]{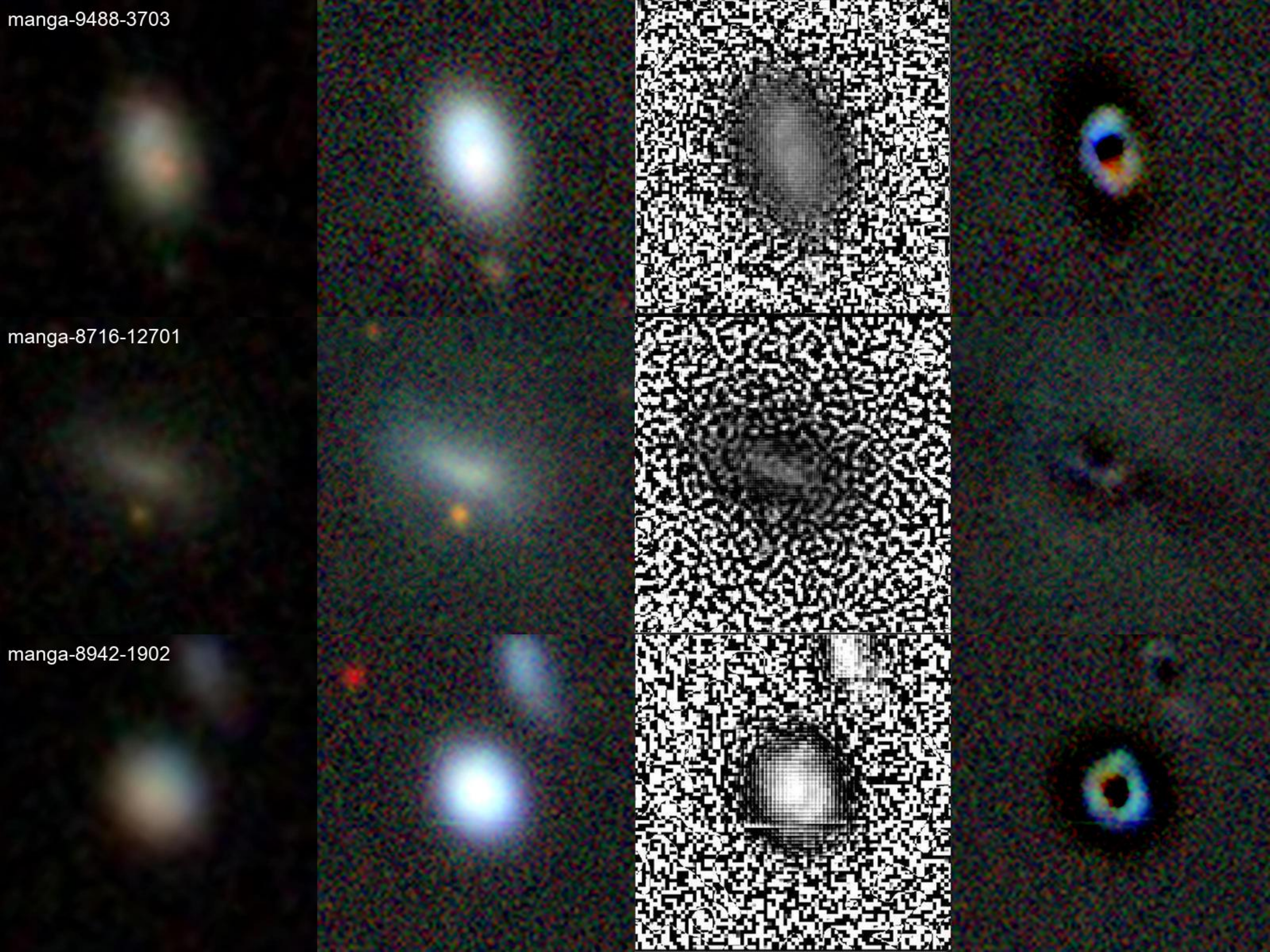}
 \caption{Galaxies manga-9488-3703, manga-8716-12701 and manga-8942-1902 are examples of galaxies difficult to classify. From left- to right-hand the \emph{gri} SDSS, the \emph{grz} DESI, the filter-enhanced and the DESI residual images. These galaxies are intrinsically faint, in the low/intermediate stellar mass range, typically concentrated and disky in nature.}
 \label{fig:faint_S}
\end{figure}

Figure~\ref{fig:faint_S} shows a few examples of galaxies in this category. Each panel shows from left to right, the SDSS $gri$ colour image, the DESI $grz$ colour image, the filter-enhanced DESI $r-$band image and the DESI residual (after best-model subtraction).

\begin{figure}
 \includegraphics[width=\columnwidth]{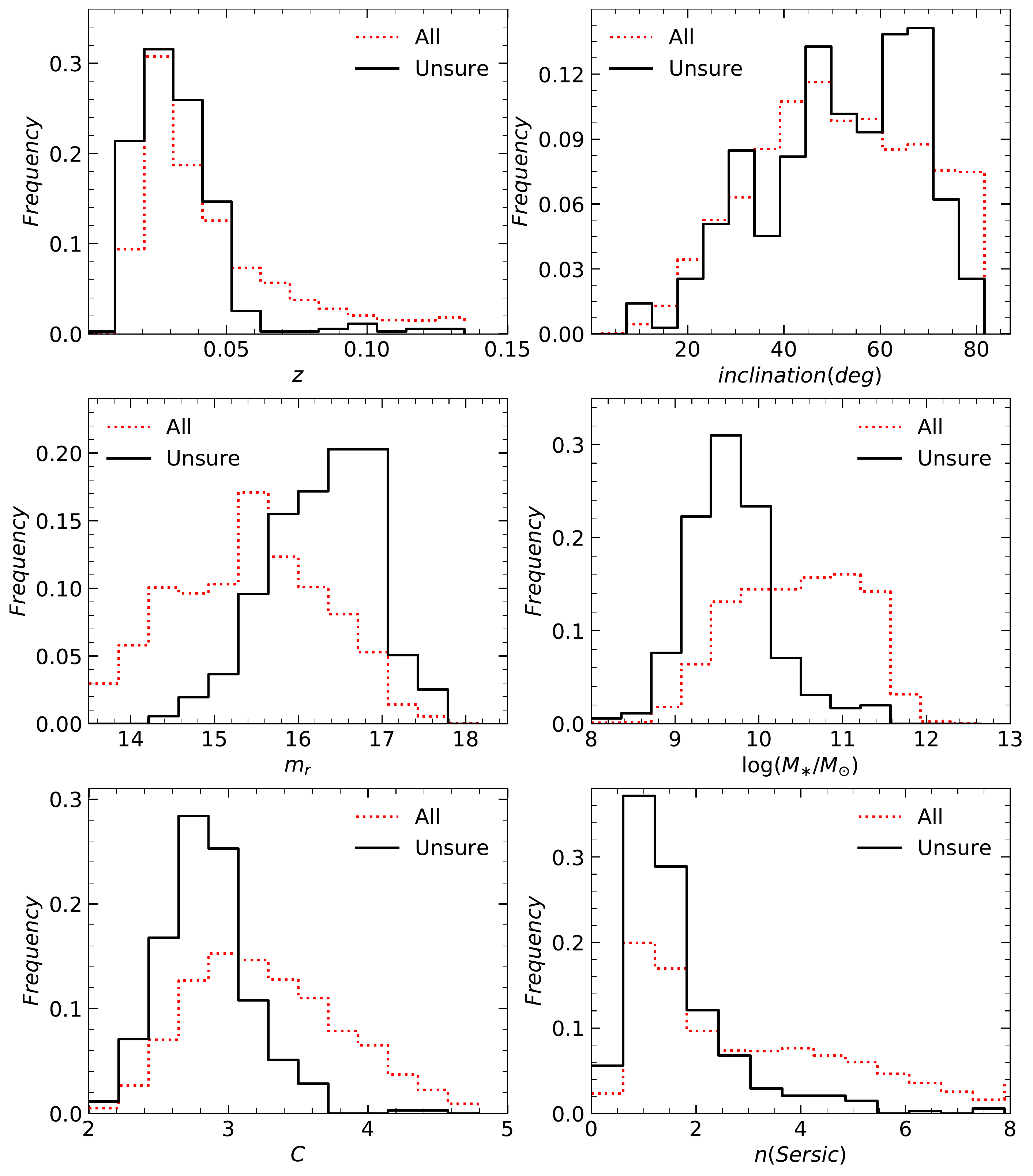}
 \caption{Frequency distribution of galaxies difficult to classify ("unsure"; solid-line) as a function of redshift, inclination, $r-$band apparent magnitude, stellar mass, concentration (C), and S\'ersic index. As reference, red dotted-lines show the distributions for the whole MaNGA DR15 sample. All these distributions are normalised to the total number in every sample. 
 }
 \label{fig:hist_dif}
\end{figure}

Figure~\ref{fig:hist_dif} presents the frequency distribution of the 355 MaNGA galaxies identified as ``unsure'' (black solid line), in terms of redshift (upper left panel), inclination (under the assumption of a measurable external disk; upper right panel), apparent $r-$band magnitude (middle left panel), stellar mass (middle right panel), concentration ($C$)\footnote{$C$ comes from our estimate to the DESI images following \citet{Conselice2003} method, reported in an upcoming paper.} (lower left panel) and S\'ersic index $n$ (lower right panel). The black dotted-line histograms correspond to the whole MaNGA sample. S\'ersic indices were taken from \citet{Fischer2019} (S\'ersic fit) after a 2D surface brightness modelling. 

Figure~\ref{fig:hist_dif} shows that most (94\%) of the MaNGA galaxies identified as ``unsure'' are in the low redshift range, with a median value $z \approx 0.028$, compared to the median $z = 0.037$ for the whole MaNGA DR15 sample.  These are visually faint (87\% having $m_r$ $>$ 15.5) with a median value $m_r \approx 16.36$, compared to a median $m_r$ = 15.44 for the whole sample. These objects are not particularly biased by inclination, but most of them are intrinsically of low masses, with a median value of $\log(\ms/\msun)=9.62$, lower than the median for the whole sample, $\log(\ms/\msun)=10.48$.
They are also low concentration galaxies with a median $C$ = 2.82 compared to  $C$ = 3.23 for the whole sample. Finally, most of these are disky with a median Sersic index $n$ = 1.33 compared to a median $n$ = 2.52 for the whole sample.     

Therefore, the MaNGA galaxies that are difficult to classify are not particularly biased by inclination and are not the most distant ones. Instead, these objects are mostly faint and of low masses, typically of low concentration and disky in nature.

\section{Conclusions}
\label{sec:conclusions}

We have carried out a detailed Hubble morphological classification for 4614 galaxies with unique ID in the MaNGA survey corresponding to SDSS DR15. For the visual inspection, we have used mosaics generated from a combination of $r-$band SDSS/DESI images, their digital image post-processing, and the  DESI residual images after subtraction of a best surface brightness model. We attempted to classify as many as possible objects, including highly inclined galaxies and galaxies in non-strongly interacting pairs (at least the main component); only 7 galaxies, clearly involved in major mergers, were impossible to classify.
On the other hand, $\approx 8\%$ of the galaxies were difficult to classify, flagging them as ``unsure''. They consist of relatively isolated galaxies mostly apparently faint and diffuse, disky in nature, and of relatively low concentration. In addition to the Hubble morphological types, we also identified the presence of bars and bright tidal features. 

The depth of the DESI images in combination with our image post-processing and DESI image by-products provided a wealth of additional morphological information in edge-on and low inclination galaxies, compared to those in the SDSS images, translating into a more refined classification. This also allowed us to achieve a better distinction between E and S0 galaxies, as well as better criteria to identify bars and tidal features.

To calculate different morphological fractions that describe the local galaxy population, we have assigned a weight to each galaxy so that the sample is complete in volume for $\ms>10^9$ \msun. Our volume completeness corrections not only depend on \ms\ and $z$, but also take into account the galaxy $g-r$ colour. The main results regarding the morphological mix in the {\it volume-corrected} MaNGA DR15 sample, including bars and tidal features, are as follows.

\begin{itemize}
    \item The frequency distribution of the morphological types is bimodal (Fig. \ref{fig:hist_morpho}), with a broad mode of LTGs from Sab to Irr types (67\%) that peaks in the Sc types, a narrow mode of ETGs (E--S0; 19\%), and a minimum between both modes in the S0a–-Sa types (14\%), which we  have called intermedium-type galaxies, ITGs.  
    
    \item The bi-variate distribution of galaxies in the \type--\ms\ diagram (Fig. \ref{fig:mass_morpho}; $i<70^\circ$) presents two densely populated regions: a broad cloud of Sab--Sd spirals of low-intermediate masses, $\log(\ms/\msun)\lesssim10.3$, and a sequence of E--S0 galaxies with masses mostly above $\log(\ms/\msun)\sim 10$. The transition zone between these two modes, the "morphological valley", is around the S0a--Sa galaxies and it is almost independent of mass. 
    
    \item The distribution of the dust attenuation-corrected $g-i$ colour in the \type--\ms\ diagram shows an approximate correspondence between the LTG cloud and blue galaxies, and between the ETG sequence and red galaxies (Fig.\ref{fig:AGE}). The above implies that the well-known bimodality in the colour--\ms\ diagram is mostly a consequence of the bimodality in the \type--\ms\ diagram. Only ITGs (S0a--Sa) present a large diversity of $g-i$ colours, from red for the largest masses to blue for the lowest masses, with a mass-dependent colour transition zone, the green valley.
    
    \item Using Pipe3D, we have calculated light-weighted ages of the MaNGA galaxies, \age. The distribution of \age\ in the \type--\ms\ diagram (Fig. \ref{fig:AGE}) is similar as the $g-i$ colour one but the former is smoother, with a sharper segregation into two populations (late-type/young and early-type/old galaxies), and a better defined transitioning zone, than the latter.  
    
    \item The fraction of barred galaxies in the MaNGA DR15 sample amounts to $\approx 47\%$.
    Barred galaxies are present in all Hubble types from S0. There is no bi-modal distribution of the the bar fraction in terms of \type, but an incipient bi-modality is observed in terms of \ms, contributed mainly by S0--Sb types on the high-mass region and by late Sbc-Sm types in the low-mass region, with a minimum at $\log(\ms/\msun)\sim$9.9 (Fig. \ref{fig:Bar_fam}). 
    In general, the fraction of barred galaxies increases at all masses from S0 to Sab types, but for later types, the fractions almost do not depend on \type, and attain values above $\sim 0.4$ (Fig. \ref{fig:Bar_massT}).  
    In terms of the $g-i$ colour, the S0--Sb barred galaxies are mostly red while the Sbc--Sm barred galaxies are mostly blue.   
    
    \item  The average fraction of galaxies with bright tidal features in the whole sample is $\approx 14\%$, with fractions of $11.4\%$ for the E--S0a types and $14.8\%$ for the Sa--Irr types.  
    In general, the tidal fraction increases in galaxies of E--Sa types as they are more massive (Figs. \ref{fig:tides_hist} and \ref{fig:Tides_massT}). The galaxies with the highest fractions of  tidal features ($\gtrsim 0.35$) are the Sdm, which are only a few and of low masses. 

\end{itemize}

We have compared the result of our visual morphological classification with that of \citetalias{Nair2010} for the 1150 galaxies in common ($z<0.1$). The agreement is in general reasonable \citep{Naim1995}, with a median scatter in \type\ of 1.2 (excluding the small fraction of Sd and later galaxies). For \type$\le$Sb, we classify on average slightly later types than \citetalias{Nair2010} for a fraction of the sample, while the opposite applies for \type$>$Sbc (Fig. \ref{fig:Histclass_compN10}). A comparison with the Machine Learning-based results of \citetalias{DominguezSanchez2021} for the same MaNGA galaxies, shows a median scatter in \type\ of 1.48, with differences in the median as a function of \type\ similar as the comparison with \citetalias{Nair2010}, but the median deviation for \type$>$Sbc is smaller when comparing with \citetalias{DominguezSanchez2021} (Fig. \ref{fig:Histclass_compF19}). 

We explain the mentioned above trends in the median deviation and scatter from the one-to-one comparisons with \citetalias{Nair2010} and \citetalias{DominguezSanchez2021} mostly as consequence of the use of deep DESI images and our image post-processing procedure, which helped to identify morphological details better in edge-on and low inclination galaxies.
We also tested our E/S0 results with an automated classification scheme by \citet{Cheng2011}, using two separation variants in the $B/T$--$b/a$--$C$ diagrams, finding a comparable distinction of E and S0 galaxies for bulge-dominated galaxies.

A comparison of our bar identification with that of \citetalias{Nair2010} for the galaxies in common ($i\le70^\circ$), shows that we recover 94\% of the barred galaxies reported in that work, but we identify a factor of $\sim 2$ more barred galaxies. A similar result is obtained when comparing with the galaxies in common with the MaNGA-GalaxyZoo catalogue. 
In particular, for types S0 and S0a we report a small but non-negligible fraction of bared galaxies compared to the almost complete absence of bars reported in those works.

We are now in the process of completing the visual morphological classification, as well as the identification of bars and tidal characteristics, for the final MaNGA sample to be released in the future (SDSS DR17). Elsewhere, we will present our results for the $\sim 10,000$ galaxies from the final MaNGA sample as a publicly available value-added catalogue. In this catalogue, we will also report the results of our parametric morphological classification based on the concentration-asymmetry-clumpiness (CAS) parameters.

\section*{Acknowledgements}

Aside from the intrinsic scientific goals in the classification of MaNGA galaxies, this was conceived as part of and Education and Public Outreach project under the supervision of the authors of this catalogue. This has been possible through a sustained participation of high school and early undergraduate students at UNAM. The MaNGA team at UNAM highlights and favors the participation of students of different school levels in Sloan IV projects.
Special thanks to: Isa\'i Chavez, Luis A. Fuentes, Brenda Garc\'ia, Juan G. Jos\'e, Aurora Le\'on, Luis C. Mascherpa, Edgar Muñoz, Nayeli Pantoja and Arlin Rodr\'iguez for their valuable contribution to image preparation and catalogue products.

JAVM was supported by the CONACyT postdoctoral fellowship program. MHE aknowledges financial suport from CONACyT 252531. ARP and VAR acknowledge financial support from CONACyT through ``Ciencia Basica'' grant 285721, and from DGAPA-UNAM through PAPIIT grant IA104118.

This research made use of Montage, funded by the National Aeronautics and Space Administration's Earth Science Technology Office, Computational Technnologies Project, under Cooperative Agreement Number NCC5-626 between NASA and the California Institute of Technology. The code is maintained by the NASA/IPAC Infrared Science Archive.
This service also uses software developed by IPAC for the US National Virtual Observatory, which is sponsored by the National Science Foundation.

We acknowledge the use of the Legacy Surveys data; full acknowledgments can be found here: \url{http://legacysurvey.org/acknowledgment/}

Funding for the Sloan Digital Sky Survey IV has been provided by the Alfred P. Sloan Foundation, the U.S. Department of Energy Office of Science, and the Participating Institutions. SDSS-IV acknowledges support and resources from the Center for High-Performance Computing at the University of Utah. The SDSS web site is www.sdss.org.

SDSS-IV is managed by the Astrophysical Research Consortium for the 
Participating Institutions of the SDSS Collaboration including the 
Brazilian Participation Group, the Carnegie Institution for Science, 
Carnegie Mellon University, the Chilean Participation Group, the French Participation Group, Harvard-Smithsonian Center for Astrophysics, 
Instituto de Astrof\'isica de Canarias, The Johns Hopkins University, 
Kavli Institute for the Physics and Mathematics of the Universe (IPMU) / 
University of Tokyo, Lawrence Berkeley National Laboratory, 
Leibniz Institut f\"ur Astrophysik Potsdam (AIP),  
Max-Planck-Institut f\"ur Astronomie (MPIA Heidelberg), 
Max-Planck-Institut f\"ur Astrophysik (MPA Garching), 
Max-Planck-Institut f\"ur Extraterrestrische Physik (MPE), 
National Astronomical Observatories of China, New Mexico State University, 
New York University, University of Notre Dame, 
Observat\'ario Nacional / MCTI, The Ohio State University, 
Pennsylvania State University, Shanghai Astronomical Observatory, 
United Kingdom Participation Group,
Universidad Nacional Aut\'onoma de M\'exico, University of Arizona, 
University of Colorado Boulder, University of Oxford, University of Portsmouth, 
University of Utah, University of Virginia, University of Washington, University of Wisconsin, 
Vanderbilt University, and Yale University.

This project makes use of the MaNGA-Pipe3D dataproducts. We thank the IA-UNAM MaNGA team for creating this catalogue, and the CONACyT-180125 project for supporting them.

\section*{DATA AVAILABILITY}

The data described in this paper has been reported as part of the MaNGA Visual Morphology catalogue DR15. This catalogue is available via the SDSS website at: \url{https://www.sdss.org/dr16/data_access/value-added-catalogs/?vac_id=manga-visual-morphologies-from-sdss-and-desi-images}




\bibliographystyle{mnras}
\bibliography{MaNGApaper} 




\appendix

\section{Tables}

\begin{landscape}
\begin{table}
  \begin{center}
    \caption{Top panel: Number of galaxies and fractions to the total sample (in parentheses) in bins of morphology and stellar mass. The lower sub-panel present the total number of galaxies and fraction, the mean stellar mass and the standard deviation of the stellar mean, for every morphological type and sub-class. Bottom panel: the volume-corrected fractions in morphology and stellar mass bins. Lower sub-panel present the total fraction, mean stellar mass and its standard deviation per morphological bin after applying volume corrections. }
    \label{tab:mass}
    \begin{tabular}{c c c c c c c c c c c c c c}
      \hline
      \multicolumn{14}{c|}{Original sample} \\
      \hline
      $\log (M_{\ast}/M_{\sun})$ & E & S0 & S0a & Sa & Sab & Sb & Sbc & Sc & Scd & Sd & Sdm & Sm & Irr \\
      \hline
$\leq$ 9 & 4 ( 0.001 ) & 0 ( 0.0 ) & 2 ( 0.0 ) & 7 ( 0.002 ) & 3 ( 0.001 ) & 4 ( 0.001 ) & 9 ( 0.002 ) & 8 ( 0.002 ) & 3 ( 0.001 ) & 3 ( 0.001 ) & 2 ( 0.0 ) & 9 ( 0.002 ) & 6 ( 0.001 ) \\ 
9 - 9.5 & 19 ( 0.004 ) & 26 ( 0.006 ) & 12 ( 0.003 ) & 31 ( 0.007 ) & 30 ( 0.007 ) & 43 ( 0.009 ) & 73 ( 0.016 ) & 93 ( 0.02 ) & 48 ( 0.01 ) & 13 ( 0.003 ) & 8 ( 0.002 ) & 11 ( 0.002 ) & 10 ( 0.002 ) \\ 
9.5 - 10 & 49 ( 0.011 ) & 101 ( 0.022 ) & 67 ( 0.015 ) & 78 ( 0.017 ) & 58 ( 0.013 ) & 136 ( 0.03 ) & 154 ( 0.033 ) & 183 ( 0.04 ) & 61 ( 0.013 ) & 13 ( 0.003 ) & 4 ( 0.001 ) & 8 ( 0.002 ) & 8 ( 0.002 ) \\ 
10 - 10.5 & 64 ( 0.014 ) & 141 ( 0.031 ) & 89 ( 0.019 ) & 86 ( 0.019 ) & 96 ( 0.021 ) & 110 ( 0.024 ) & 158 ( 0.034 ) & 141 ( 0.031 ) & 33 ( 0.007 ) & 7 ( 0.002 ) & 4 ( 0.001 ) & 1 ( 0.0 ) & 1 ( 0.0 ) \\ 
10.5 - 11 & 119 ( 0.026 ) & 172 ( 0.037 ) & 82 ( 0.018 ) & 131 ( 0.028 ) & 127 ( 0.028 ) & 154 ( 0.033 ) & 120 ( 0.026 ) & 99 ( 0.021 ) & 14 ( 0.003 ) & 1 ( 0.0 ) & 1 ( 0.0 ) & 0 ( 0.0 ) & 1 ( 0.0 ) \\ 
11 - 11.5 & 420 ( 0.091 ) & 156 ( 0.034 ) & 47 ( 0.01 ) & 79 ( 0.017 ) & 92 ( 0.02 ) & 86 ( 0.019 ) & 64 ( 0.014 ) & 33 ( 0.007 ) & 5 ( 0.001 ) & 0 ( 0.0 ) & 0 ( 0.0 ) & 0 ( 0.0 ) & 0 ( 0.0 ) \\ 
11.5 - 12 & 176 ( 0.038 ) & 54 ( 0.012 ) & 5 ( 0.001 ) & 5 ( 0.001 ) & 10 ( 0.002 ) & 6 ( 0.001 ) & 8 ( 0.002 ) & 0 ( 0.0 ) & 0 ( 0.0 ) & 0 ( 0.0 ) & 0 ( 0.0 ) & 0 ( 0.0 ) & 0 ( 0.0 ) \\ 
$\geq$ 12 & 3 ( 0.001 ) & 3 ( 0.001 ) & 0 ( 0.0 ) & 0 ( 0.0 ) & 1 ( 0.0 ) & 2 ( 0.0 ) & 1 ( 0.0 ) & 1 ( 0.0 ) & 0 ( 0.0 ) & 0 ( 0.0 ) & 0 ( 0.0 ) & 0 ( 0.0 ) & 0 ( 0.0 ) \\ 
\hline
N gal & 854 ( 0.185 ) & 653 ( 0.142 ) & 304 ( 0.066 ) & 417 ( 0.091 ) & 417 ( 0.091 ) & 541 ( 0.117 ) & 587 ( 0.127 ) & 558 ( 0.121 ) & 164 ( 0.036 ) & 37 ( 0.008 ) & 19 ( 0.004 ) & 29 ( 0.006 ) & 26 ( 0.006 ) \\ 
$\langle \log M_{\ast} \rangle$ & 11.07 & 10.63 & 10.41 & 10.41 & 10.5 & 10.36 & 10.21 & 10.04 & 9.82 & 9.6 & 9.58 & 9.21 & 9.22 \\ 
$\sigma \langle \log M_{\ast} \rangle$ & 0.6 & 0.65 & 0.58 & 0.65 & 0.64 & 0.64 & 0.64 & 0.59 & 0.53 & 0.47 & 0.52 & 0.47 & 0.71 \\ 

    \hline
      \multicolumn{14}{c|}{Volume-corrected fractions} \\
    \hline
      $\log (M_{\ast}/M_{\sun})$ & E & S0 & S0a & Sa & Sab & Sb & Sbc & Sc & Scd & Sd & Sdm & Sm & Irr \\
      \hline
9 - 9.5 & 0.013 & 0.025 & 0.009 & 0.026 & 0.023 & 0.042 & 0.06 & 0.075 & 0.036 & 0.012 & 0.01 & 0.012 & 0.012 \\
9.5 - 10 & 0.013 & 0.03 & 0.016 & 0.02 & 0.018 & 0.037 & 0.044 & 0.051 & 0.02 & 0.005 & 0.001 & 0.003 & 0.003 \\
10 - 10.5 & 0.014 & 0.035 & 0.017 & 0.02 & 0.024 & 0.025 & 0.038 & 0.038 & 0.009 & 0.003 & 0.001 & 0.0 & 0.0 \\
10.5 - 11 & 0.013 & 0.024 & 0.01 & 0.018 & 0.015 & 0.017 & 0.012 & 0.015 & 0.001 & 0.0 & 0.0 & 0.0 & 0.0 \\
11 - 11.5 & 0.014 & 0.005 & 0.001 & 0.003 & 0.003 & 0.002 & 0.002 & 0.002 & 0.0 & 0.0 & 0.0 & 0.0 & 0.0 \\
11.5 - 12 & 0.002 & 0.0 & 0.0 & 0.0 & 0.0 & 0.0 & 0.0 & 0.0 & 0.0 & 0.0 & 0.0 & 0.0 & 0.0 \\
$\geq$ 12 & 0.001 & 0.0 & 0.0 & 0.0 & 0.0 & 0.0 & 0.0 & 0.0 & 0.0 & 0.0 & 0.0 & 0.0 & 0.0 \\
\hline
Total frac & 0.07 & 0.12 & 0.054 & 0.087 & 0.084 & 0.122 & 0.156 & 0.18 & 0.066 & 0.02 & 0.012 & 0.015 & 0.015 \\
$\langle \log(M_{\ast}/M_{\sun}) \rangle$ & 10.34 & 10.05 & 10.06 & 9.94 & 9.96 & 9.84 & 9.76 & 9.74 & 9.54 & 9.48 & 9.33 & 9.27 & 9.21 \\
$\sigma (\log(M_{\ast}/M_{\sun})$ & 0.73 & 0.6 & 0.52 & 0.61 & 0.61 & 0.56 & 0.51 & 0.52 & 0.43 & 0.38 & 0.35 & 0.22 & 0.24 \\

      \hline
     \multicolumn{14}{l|}{NOTE: Total number of galaxies and fractions in each column may not be the same to those reported in Table~\ref{tab:morpho}, due to the lack of information of stellar mass for 8 galaxies. } \\
    \end{tabular}
  \end{center}
\end{table}
\end{landscape}

\bsp	
\label{lastpage}
\end{document}